\documentclass[pra,twocolumn,amsmath,amssymb,superscriptaddress]{revtex4-1}

\usepackage{epsfig,amsmath}
\usepackage{subfigure}
\usepackage{graphicx}
\usepackage{dcolumn}
\usepackage{stmaryrd}
\usepackage{mathrsfs}
\usepackage{pifont}
\usepackage{amsthm}
\usepackage{amssymb}
\usepackage{bm}
\usepackage{latexsym}
\usepackage[colorlinks=true,linkcolor=blue,citecolor=blue]{hyperref}
\usepackage{color}
\usepackage{epstopdf}

\begin{document}

\title{Magnon-assisted photon-phonon conversion in the presence of the structured environments}

\author{Shi-fan Qi}
\affiliation{Zhejiang Province Key Laboratory of Quantum Technology and Device, Department of Physics, Zhejiang University, Hangzhou 310027, Zhejiang, China}

\author{Jun Jing}
\email{Email address: jingjun@zju.edu.cn}
\affiliation{Zhejiang Province Key Laboratory of Quantum Technology and Device, Department of Physics, Zhejiang University, Hangzhou 310027, Zhejiang, China}

\date{\today}

\begin{abstract}
Quantum conversion or interface is one of the most prominent protocols in quantum information processing and quantum state engineering. We propose a photon-phonon conversion protocol in a hybrid magnomechanical system comprising a microwave optical mode, a driven magnon mode and a mechanical-vibrating mode, which has attracted much interest and is expected to become a building block of the future quantum information network due to its controllability in coupling strengths. The microwave photons in the optical cavity are coupled to the magnons by the Zeeman interaction, and the latter are coupled to the mechanical phonons by the magnetostrictive interaction. With strong photon-magnon interaction and strong driving on magnon, an effective Hamiltonian is constructed to describe the conversion between photons and phonons nearby their resonant point. The cavity-magnon system can then play the role of a quantum memory. Moreover, the faithfulness of the photon-phonon conversion is estimated in terms of fidelities for state evolution and state-independent transfer. The former is discussed in the Lindblad master equation taking account the leakages of photon, phonon and magnon into consideration. The latter is derived by the Heisenberg-Langevin equation considering the non-Markovian noise from the structured environments for both optical and mechanical modes. The state-evolution fidelity is found to be robust to the weak leakage. The transfer fidelity can be maintained by the Ohmic and sub-Ohmic environments of the photons and is insensitive to the $1/f$ noise of the phonons. Our work thus provides an interesting and novel application for the magnon system as a photon-phonon converter in the microwave regime.
\end{abstract}

\maketitle
\section{Introduction}

The cavity-magnon system is a rapid-developing mesoscopic platform for quantum information processing~\cite{magnon,magnon2,magnon3,magnon4}. As an active topic of research, it has been theoretically investigated~\cite{yigcavity1,yigcavity2,yigcavity3,magnonblockade} and experimentally demonstrated~\cite{yigcavity3,yigcavity4,yigcavity5,yigcavity6,magnonqubit,magnonqubit2} in the perspective of cavity quantum electrodynamics (cQED) during the past decade. Following the circuit-QED~\cite{superconducting} systems and the semiconductor microcavities systems (such as quantum dots embedded in cavities~\cite{quantumdot,quantumdot2}), the cavity-magnon system actually becomes an alternative candidate to exploit the ubiquitous effects of cQED in the strong-coupling regime~\cite{strongcoupling1,strongcoupling2,strongcoupling3}.

Typically in a cavity-magnon system, a magnet-spin ensemble in a single-crystal yttrium iron garnet (YIG) sphere is loaded into a high-Q cavity. Down to the quantum level, it is found that the Kittel mode of the spin ensemble (the ground state) in the YIG sphere could be strongly coupled to the microwave photons in a cavity-mode, and in the mean time the magnons are coupled to the phonons describing the mechanical vibration of the same sphere (See the diagram in Fig.~\ref{model}). Then a cavity-magnon system could be considered as a cavity-magnomechanical system taking inspiration from the cavity optomechanics, where the optical mode is {\em directly} coupled to the mechanical vibration of the movable mirror via radiation pressure~\cite{optcavity}. Significant progresses regarding the mechanical vibration have been reported to demonstrate the quantum advantages, such as the phonon cooling in a non-Markovian environment~\cite{optcool}, the quantum entanglement between mechanical elements and cavity modes~\cite{optentang1,optentang2}, and the quantum state transfer between photon and phonon~\cite{opttransfer}. In light of displaying the quantum characteristics, it is interesting to develop original ideas in the existing hybrid quantum systems~\cite{qed,qed2,qed3,qed4,optcavity} to the cavity-magnomechanical system. For example, the first implementation of the photon-magnon-phonon interaction was addressed in Ref.~\cite{magnoncavity}, based on which the photon-magnon-phonon entanglement in cavity magnomechanics has been proposed in Ref.~\cite{mppentang}.

In terms of quantum conversion, protocols for light-matter interface and state-transfer have been implemented with atomic systems under cavity-QED~\cite{statetransfer1,statetransfer2}. The scheme about quantum state conversion between microwave and optical photons was proposed via an optoelectromechanical interface~\cite{stateconversion}. Optomechanical systems can also serve as a light-matter interface, in which quantum information and quantum fluctuations originally encoded in an optical field can be reversibly mapped to a mechanical oscillator with a much smaller decay rate~\cite{statetransfer3,statetransfer4}, by which the mechanical oscillator serves as a quantum network node~\cite{magnoncavity,mppentang} for potential information storage and processing~\cite{stateprocess}. In this work, we present a protocol for photon-phonon conversion in the cavity magnomechanical system~\cite{mppentang} on account of the fundamental interest in the physical process, also exploiting the strong coupling for the magnon-photon and the magnon-phonon interactions and the negligible decay rate of the phonon. Note the phonon in YIG sphere decays with a rate about $100$ Hz~\cite{magnoncavity}, much smaller than its own frequency and that of the photons. In comparison to the direct photon-phonon interaction in the well-studied cavity optomechanical system, our proposal realizes indirect coupling photon-phonon interaction with the same order of magnitude in a three-mode system. A more important advantage of the cavity-magnomechanical system~\cite{magnoncavity} is its potential large-interval parametric implementation due to the fact that the single coupling strength between magnon and phonon can be tuned from $0$ to $60$ MHz by manipulating the direction of the bias fields.

The strong coupling inside a hybrid system opens a door to study the physics of virtual processes governed by the interaction Hamiltonian and leads to many interesting phenomena and applications in quantum engineering~\cite{strongcoupling1,strongcoupling2,strongcoupling3}. For example, the interesting Bell states, GHZ states~\cite{state1} and NOON states~\cite{state2} can be generated in the strongly coupled circuit-QED systems by virtue of the multiple-photon process. We find the core idea for these general Rabi models~\cite{rabi2,state1,state2} describing the a single atomic system strongly coupled to a harmonic oscillator system can be used to construct an effective Hamiltonian describing the indirect coupling between photon and phonon at their resonant point via the virtual process of magnons. In particular, the effective coupling obtained by the second-order perturbation theory~\cite{rabi2,state2} supports a novel conversion protocol between the mechanical-oscillation and the microwave optical mode, resembling the conventional one in the optomechanical systems.

The rest part of this work is structured as following. In Sec.~\ref{system}, we introduce the hybrid quantum model and derive the effective Hamiltonian for the photon-phonon conversion mediated by the magnon mode. In Sec.~\ref{fidelity}, the effective Hamiltonian is confirmed by comparing to the original Hamiltonian with respect to the effective coupling strength and is phenomenologically tested in terms of the state evolution under the leakage of photon, magnon and phonon by a Lindblad master equation. While in Sec.~\ref{dynamics}, a state-independent transfer fidelity determined by the effective Hamiltonian is derived through a microscopic analysis via the Heisenberg-Langevin equation. It is shown that our conversion protocol can be implemented with a high fidelity under certain structured environments for the photon mode. In Sec.~\ref{diss}, we discuss the Kerr effect about the magnon mode on the effective Hamiltonian and then summarize the whole work.

\section{Model and the effective Hamiltonian}\label{system}

\begin{figure}[htbp]
\centering
\includegraphics[width=0.35\textwidth]{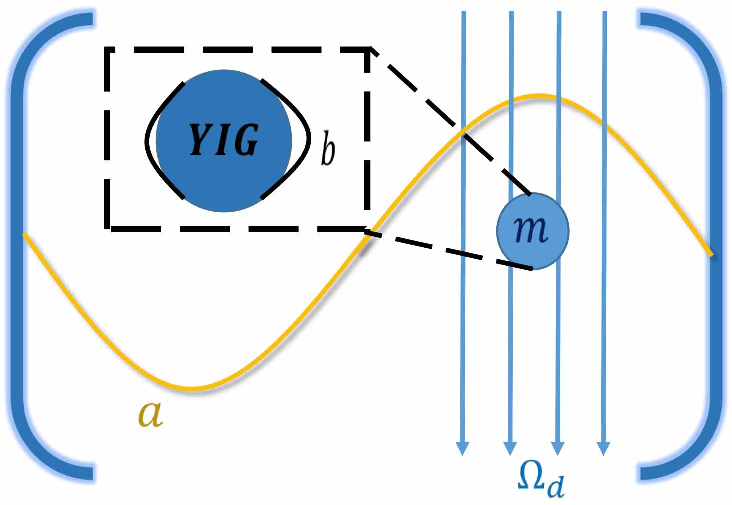}
\caption{(Color online) Model diagram: a YIG sphere is placed inside a microwave cavity near the maximum magnetic field of the cavity mode, and simultaneously in a uniform bias magnetic field, which establishes the magnon-photon coupling. The magnon mode is driven by a microwave source (with a magnitude $\Omega_d$) to enhance the magnonmechanical coupling.}\label{model}
\end{figure}

We consider a hybrid cavity magnonmechanical system as shown in Fig.~\ref{model}, which is constituted by cavity-mode photons, magnons, and phonons~\cite{mppentang} down to the quantum level. The magnons are embodied by a collective motion of a large number of spins in a ferrimagnet, e.g., a YIG sphere. The magnons are coupled to the cavity photons via Zeeman interaction of magnetic
field of light and magnetization of the magnet. The coupling between magnons and phonons is mediated by the magnetostrictive interaction. In particular, the temporal-varying magnetization induced by the magnon excitation inside the YIG sphere leads to the deformation of its geometrical structure, which forms the vibrational modes (phonons) of the sphere. The sphere size is considered to be much smaller than the wavelength of the microwave photons, such that the magnetic dipole interactions are isotropic and the effect of radiation pressure (proportional to the photon number $\langle a^\dag a\rangle$) over the sphere is negligible. The system Hamiltonian thus reads ($\hbar\equiv1$)
\begin{equation}\label{Hs}
\begin{aligned}
H_S&=\omega_a a^{\dag}a+\omega_m m^{\dag}m+\omega_b b^{\dag}b\\
&+i\Omega_d(m^\dag e^{-i\omega_dt}-me^{i\omega_dt})\\
&+g_{ma}(am^\dag+a^\dag m)+g_{mb}m^{\dag}m(b+b^\dag).
\end{aligned}
\end{equation}
Here $a(a^\dag)$, $m(m^\dag)$ and $b(b^{\dag})$ are the annihilation (creation) operators of the photon, magnon and phonon modes, respectively. $\omega_a$, $\omega_m$ and $\omega_b$ are their respective eigen-frequencies. The Rabi frequency $\Omega_d=\gamma\sqrt{5N}B_0/4$ describes the coupling strength between the driving field with amplitude $B_0$ and frequency $\omega_d$ and the magnon mode, where $\gamma$ is the gyromagnetic ratio and $N$ is the number of spins~\cite{mppentang}. The magnet-spin ensemble has good coherence properties~\cite{magnoncoherence} and strong dipole transitions for efficient coupling to the microwave photons. Thus the magnon-microwave coupling strength $g_{ma}$ can be larger than the dissipation rates of both cavity and magnon modes, $\kappa_a$ and $\kappa_m$, entering into the strong-coupling regime, $g_{ma}>\kappa_a,\kappa_m$. In the Hamiltonian of Eq.~(\ref{Hs}), the photon-magnon interaction term is a result of linearizing magnetization used Holstein-Primakoff transformation and assuming that the photon magnetic field is of linear polarization and perpendicular to the static magnetization. And we have employed the rotating-wave approximation, i.e., $g_{ma}(a+a^\dag)(m+m^\dag)\to g_{ma}(am^\dag+a^\dag m)$, that is valid when $\omega_a,\omega_m\gg g_{ma}$~\cite{magnoncavity}. The single-magnon magnomechanical coupling strength $g_{mb}$ is typically small considering the large frequency mismatch between the magnon and the phonon modes, yet it can be compensated by a strong parametric drive $\Omega_d$. In this case, the magnomechanical coupling is described by a radiation pressure-like, dispersive interaction Hamiltonian $g_{mb}m^{\dag}m(b+b^\dag)$.

It is convenient to change the description of the microwave photon mode and the magnon mode by switching to a frame rotating at the driving frequency $\omega_d$. Applying the unitary transformation $U=\exp(i\omega_dta^\dag a+i\omega_dt m^\dag m)$ makes the driving terms time independent~\cite{optcavity}, and generates a rotating Hamiltonian $H_S'=UHU^\dag-iU\partial U^\dag/\partial t$ of the form
\begin{equation}\label{Hsrot}
\begin{aligned}
H_S'&=\Delta_aa^{\dag}a+\Delta_m m^{\dag}m+\omega_b b^{\dag}b+g_{mb}m^{\dag}m(b+b^\dag)\\
&+g_{ma}(am^\dag+a^\dag m)+i\Omega_d(m^\dag-m),
\end{aligned}
\end{equation}
where $\Delta_a\equiv\omega_a-\omega_d$ and $\Delta_m\equiv\omega_m-\omega_d$. The magnon mode under driving is assumed to have a large expectation value $|\langle m\rangle|\equiv M\gg1$, which allows to linearize~\cite{optcavity,opttransfer2} the system dynamics by writing the operator $m=M+\delta m$ with $\delta m$ describing the fluctuation of the magnon mode. The Hamiltonian $H_S'$ turns out to be
\begin{equation}\label{H}
\begin{aligned}
H&=H_0+V, \\
H_0&=\Delta_aa^{\dag}a+\Delta_m \delta m^{\dag}\delta m+\omega_b b^{\dag}b, \\
V&=G(\delta m^{\dag}+\delta m)(b+b^\dag)+g(a\delta m^\dag+a^\dag \delta m),
\end{aligned}
\end{equation}
where $g=g_{ma}$, and $G=M g_{mb}$ is the effective magnomechanical coupling strength. During the derivation over Eq.~(\ref{H}), all the linear terms have been omitted by an appropriate modulation over the detunings, since they indicate the presence of the average displacements. Then we only keep the quadratic interaction $V$~\cite{optcavity}. For simplicity, we apply the convention $\delta m\rightarrow m$ in the following content.

To realize the photon-phonon conversion assisted by the magnon mode via the linearized approximate Hamiltonian in Eq.~(\ref{H}), generally one can extract an effective transition from the near-degenerate subspaces based on the standard perturbation theory with respect to the coupling strengths $g$ and $G$. When the phonon-frequency $\omega_b$ is near-resonant with the detuning frequency $\Delta_a$, and both $\omega_b$ and $\Delta_a$ are far-off-resonant from the detuning frequency $\Delta_m$, i.e., $\omega_b\approx\Delta_a\gg g,G$ and $|\omega_b-\Delta_m|\gg g,G$, it is found that the tensor-product state $|n_al_mk_b\rangle\equiv|n\rangle_a|l\rangle_m|k\rangle_b$ is near-degenerate with $|(n-1)_al_m(k+1)_b\rangle$. Here the subscripts $a,m,b$ respectively represent the photon, magnon and phonon modes, and $n,l,k$ indicate their individual Fock-states.

\begin{figure}[htbp]
\centering
\includegraphics[width=0.4\textwidth]{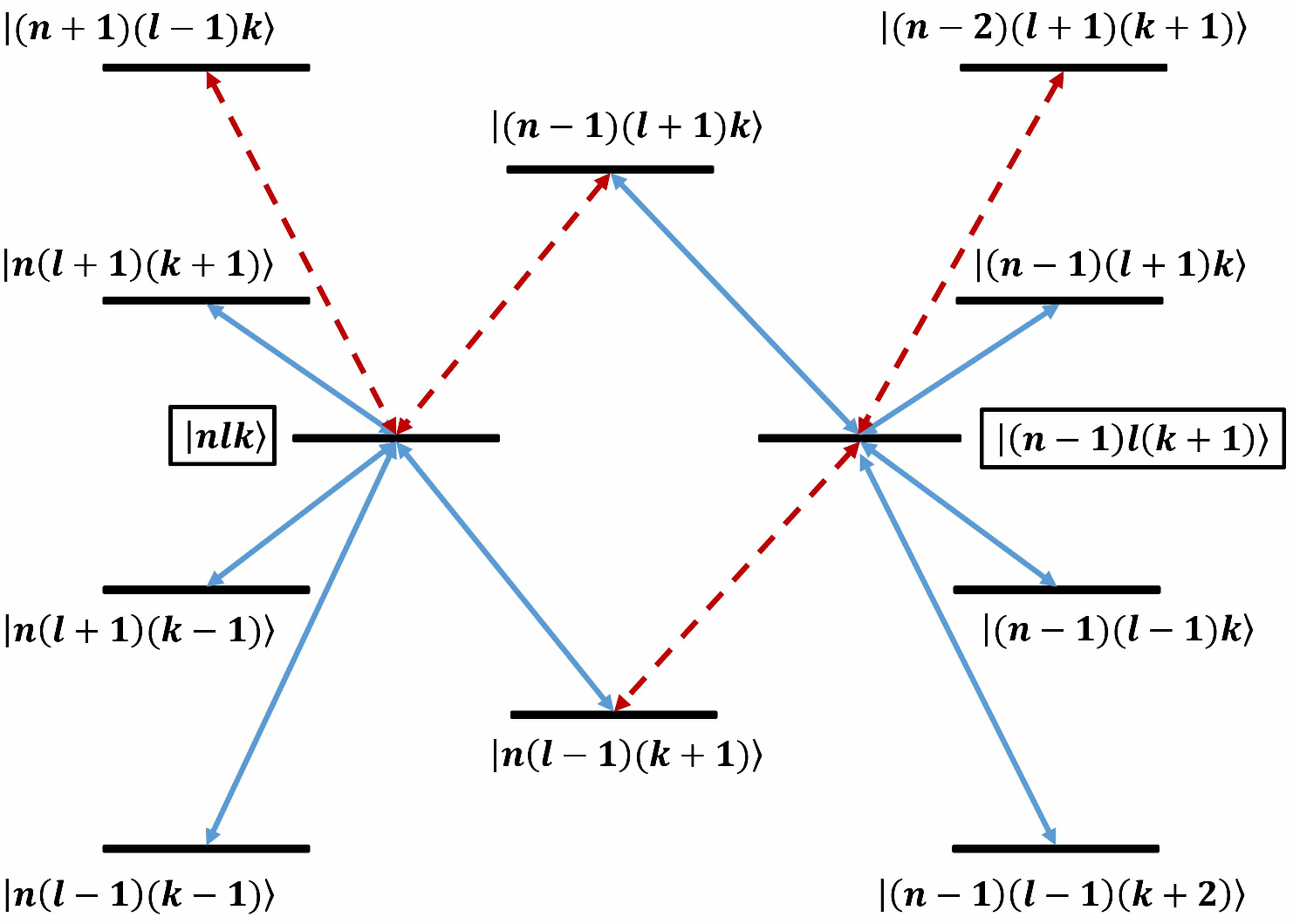}
\caption{(Color online) All the second-order (leading-order) paths involving arbitrary base-pair $|nlk\rangle\equiv|n\rangle_a|l\rangle_m|k\rangle_b$ and $|(n-1)l(k+1)\rangle$. Blue solid lines mark the transitions mediated by the photon-magnon coupling. Red long-dashed lines mark the transitions mediated by the magnon-phonon coupling. }\label{path}
\end{figure}

The indirect connection for any two eigenstates $|i\rangle$ and $|j\rangle$ of the unperturbed Hamiltonian $H_0$ can be constructed to the leading order by~\cite{rabi2,state1}
\begin{equation}\label{secondp}
g_{\rm eff}=\sum_{n\neq i,j}\frac{V_{jn}V_{ni}}{\omega_i-\omega_n},
\end{equation}
where $V_{nm}\equiv\langle n|V|m\rangle$ and $\omega_n$ is the eigenenergy of state $|n\rangle$, provided the interaction Hamiltonian $V$ is regarded as a perturbation to $H_0$.

A good approximation of the effective Hamiltonian describing the transition between arbitrary base-pair $|nlk\rangle$ and $|(n-1)l(k+1)\rangle$ can be analytically obtained using the preceding second-order perturbation theory. It can be expressed in the form
\begin{equation}\label{He12}
\begin{aligned}
H_{\rm eff}&=(\Delta_{a}+\epsilon_1)|nlk\rangle\langle nlk|\\
&+(\omega_b+\epsilon_2)|(n-1)l(k+1)\rangle\langle(n-1)l(k+1)| \\
&+g_{\rm eff}(|nlk\rangle\langle (n-1)l(k+1)|+h.c.),
\end{aligned}
\end{equation}
where $\epsilon_1$ and $\epsilon_2$ are the energy shifts due to the effective coupling for the states $|nlk\rangle$ and $|(n-1)l(k+1)\rangle$, respectively, and $g_{\rm eff}$ is the effective coupling strength. These are three coefficients to be determined in this ansatz.

We first consider the energy shift $\epsilon_1$ for the state $|nlk\rangle$. Summarizing all the 6 paths from $|nlk\rangle$ and back to itself through an intermediate state (see Fig.~\ref{path}), e.g., $|nlk\rangle\to|n(l-1)(k-1)\rangle\to|nlk\rangle$, one can obtain the second-order energy correction (shift) $\epsilon_1$ according to Eq.~(\ref{secondp})
\begin{equation}\label{g11}
\begin{aligned}
\epsilon_1&=\frac{(n-l)g^2}{\Delta_a-\Delta_m}+\frac{(k-l)G^2}{\omega_b-\Delta_m}-\frac{(l+k+1)G^2}{\Delta_m+\omega_b}.
\end{aligned}
\end{equation}
Similarly the energy shift $\epsilon_2$ for the state $|(n-1)l(k+1)\rangle$ is found to be
\begin{equation}\label{g22}
\begin{aligned}
\epsilon_2&=\frac{(n-l-1)g^2}{\Delta_a-\Delta_m}+\frac{(k-l+1)G^2}{\omega_b-\Delta_m}-\frac{(m+l+2)G^2}{\Delta_m+\omega_b}.
\end{aligned}
\end{equation}

Note an exact photon-phonon resonance facilitated by Eq.~(\ref{He12}) allows a complete Rabi oscillation between arbitrary $|nlk\rangle$ and $|(n-1)l(k+1)\rangle$, which requires that the first two terms in Eq.~(\ref{He12}) constitute the identity operator in the relevant subspace. Thus $\Delta_a+\epsilon_1=\omega_b+\epsilon_2$. Assuming the distance between $\Delta_a$ and $\omega_b$ is $\delta$, one can then have
\begin{equation}\label{deltap}
\begin{aligned}
\delta&\equiv\Delta_a-\omega_b=\epsilon_2-\epsilon_1\\
&=\frac{G^2}{\omega_b-\Delta_m}-\frac{g^2}{\Delta_a-\Delta_m}-\frac{G^2}{\omega_b+\Delta_m}\\
&=\frac{G^2-g^2}{\omega_b-\Delta_m}-\frac{G^2}{\omega_b+\Delta_m}-\frac{g^2}{(\omega_b-\Delta_m)^2}\delta+\mathcal{O}(\delta^2)\\
&=A-B\delta+\mathcal{O}(\delta^2),
\end{aligned}
\end{equation}
where $A\equiv(G^2-g^2)/(\omega_b-\Delta_m)-G^2/(\omega_b+\Delta_m)$, $B\equiv g^2/(\omega_b-\Delta_m)^2$, and $\mathcal{O}(\delta^2)$ represents all the higher orders of $\delta$ from the first order in Taylor expansion. Then $\delta$ is consistently solved as $\delta=A/(1+B)$ up to the second-order correction. Note $B\approx\mathcal{O}(g^2/|\omega_b-\Delta_m|^2)$, so that up to the second order of the coupling strengths $g$ or $G$, we have
\begin{equation}\label{delta}
\delta=\frac{G^2-g^2}{\omega_b-\Delta_m}-\frac{G^2}{\omega_b+\Delta_m}.
\end{equation}
Note $\delta$ is a Fock-state-independent coefficient in comparison to both $\epsilon_1$ and $\epsilon_2$.

Next we consider the contribution from the two paths connecting $|nlk\rangle$ and $|(n-1)l(k+1)\rangle$ in Fig.~\ref{path}, i.e., $|nlk\rangle\to|(n-1)(l+1)k\rangle\to|(n-1)l(k+1)\rangle$ and $|nlk\rangle\to|n(l-1)(k+1)\rangle\to|(n-1)l(k+1)\rangle$, to their effective coupling strength. By virtue of Eq.~(\ref{secondp}), one can have
\begin{equation}\label{g12}
\begin{aligned}
g_{\rm eff}&=\frac{(l+1)\sqrt{n(k+1)}Gg}{\Delta_a-\Delta_m}-\frac{l\sqrt{n(k+1)}Gg}{\omega_b-\Delta_m}\\
&=\frac{(l+1)\sqrt{n(k+1)}Gg}{\omega_b-\Delta_m}\left[1-\frac{\delta}{\omega_b-\Delta_m}+\mathcal{O}(\delta^2)\right]\\
&-\frac{l\sqrt{n(k+1)}Gg}{\omega_b-\Delta_m}\\
&=\sqrt{n(k+1)}\frac{Gg}{\omega_b-\Delta_m},
\end{aligned}
\end{equation}
up to the second order of the coupling strengths $g$ or $G$. Eventually the effective Hamiltonian in Eq.~\eqref{He12} can be written as
\begin{equation}\label{Heff12}
\begin{aligned}
H_{\rm eff}^{(nk)}&=g_{\rm eff}\left(|nlk\rangle\langle(n-1)l(k+1)|+h.c.\right)\\
&=g_{\rm eff}\left(|nk\rangle\langle(n-1)(k+1)|+h.c.\right)|\otimes|l\rangle\langle l|.
\end{aligned}
\end{equation}
The effective Hamiltonian extended to the whole Hilbert space of photon and phonon is therefore found to be
\begin{equation}\label{Heff}
H_{\rm eff}=\tilde{G}(ab^\dag+ba^\dag).
\end{equation}
Here
\begin{equation}\label{G}
\tilde{G}=\frac{Gg}{\omega_b-\Delta_m}
\end{equation}
is the effective coupling strength for the two modes, which is in the same order as the deviation $\delta$ of the avoided level-crossing point from $\omega_b$. More importantly, both $\tilde{G}$ and $\delta$ are Fock-state independent and robust to the variation in the magnon occupation-number, which is the key to a generic conversion rather than the state transfer in special subspaces. It should be emphasised that the same effective Hamiltonian could also be obtained by adiabatic elimination~\cite{ae1,ae2,ae3} in analog hybrid systems. The adiabatic-elimination technique forces the steady value of the fast variables into the dynamical equation to inversely obtain the effective Hamiltonian and then calculate the system shift $\delta$ and the effective coupling strength $g_{\rm eff}$ through some optimization procedure in the interested subspace. While the preceding treatment provides an alternative bottom-up approach to firstly derive these parameters and then obtain the effective Hamiltonian.

\begin{figure}[htbp]
\centering
\includegraphics[width=0.4\textwidth]{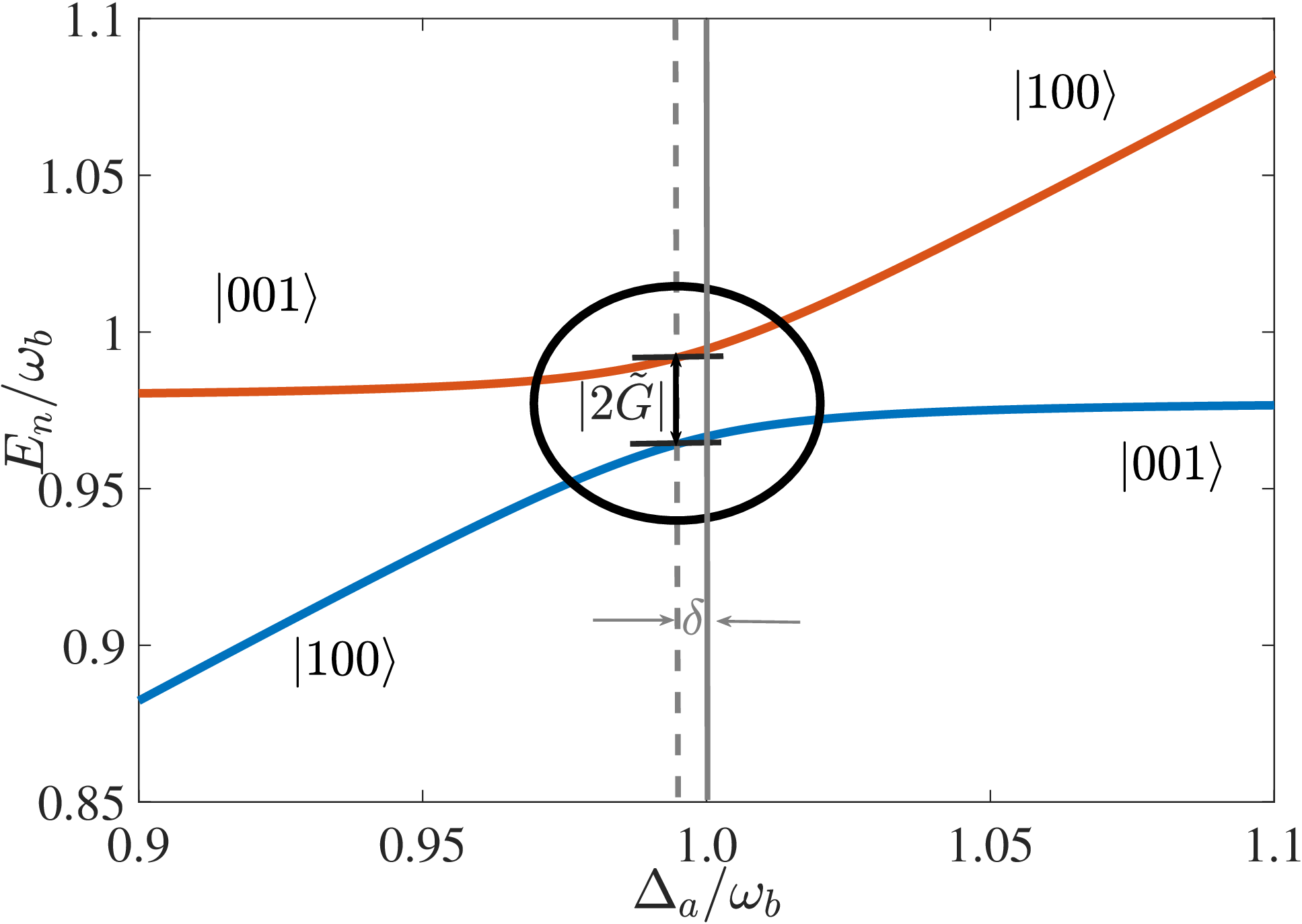}
\caption{(Color online) Normalized energy-level diagram and avoided level-crossing for the photon-phonon conversion in a single-exciton subspace, which are plotted as a function of the detuning frequency $\Delta_a/\omega_b$. Apparently the interactions in the system shift the avoided-level-crossing point from $\Delta_a/\omega_b=1$ by $\delta$ as given by Eq.~(\ref{delta}). Here we fix $\Delta_m=1.7\omega_b$ and $g=G=0.1\omega_b$. }\label{eigen1}
\end{figure}

The effective Hamiltonian in Eq.~(\ref{Heff}) is conserved in the excitation number, so that it can be written in a block-diagonal matrix formation on the Fock-state basis. For example, if we focus on the single-exciton subspace $\{|001\rangle\equiv|0\rangle_a|0\rangle_m|1\rangle_b,|100\rangle\}$ in which the magnon remains at the ground state, then we can have an $X$-gate:
\begin{equation}\label{Heff1}
H_{\rm eff}^{(1)}=\begin{bmatrix}
0 & \tilde{G}\\
\tilde{G} & 0
\end{bmatrix}=\tilde{G}\sigma_x.
\end{equation}
We plot the associated energy levels in Fig.~\ref{eigen1}, where the eigenvalues $\{E_n\}$ and the eigenstates of the Hamiltonian in Eq.~(\ref{H}) are obtained by the standard numerical diagonalization method in a truncated Hilbert space. An avoided level-crossing (distinguished in the dark circle) presents between two eigenstates of the original Hamiltonian in Eq.~(\ref{H}), when the detuning frequency of photon $\Delta_a$ approaches (but not exactly equals) the frequency of phonon $\omega_b$. The mutual interaction between the photon and the magnon and that between the magnon and the phonon induce a negative shift from $\Delta_a=\omega_b$. Here we choose $\Delta_m=1.7\omega_b$ to avoid unnecessary exciting of the magnon mode. When the system is prepared as $|100\rangle$, it can be converted to $|001\rangle$ via a half Rabi oscillation determined by $H_{\rm eff}^{(1)}$ after a duration $\pi/|2\tilde{G}|$.

\begin{figure}[htbp]
\centering
\includegraphics[width=0.4\textwidth]{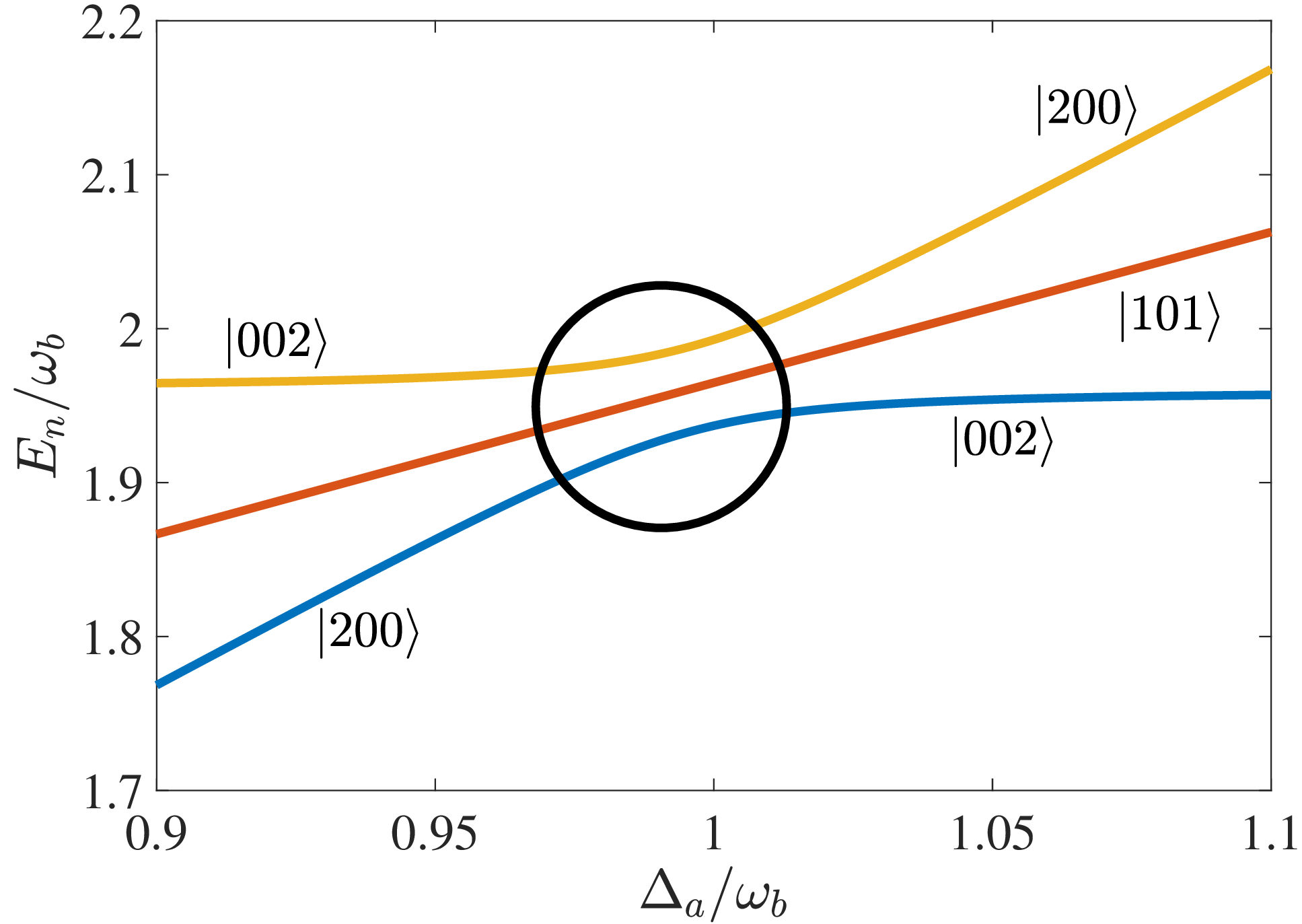}
\caption{(Color online) Normalized energy-level diagram and avoided level-crossing in a double-exciton subspace, which are plotted as a function of the detuning frequency $\Delta_a/\omega_b$. Here we fix $\Delta_m=1.7\omega_b$ and $g=G=0.1\omega_b$. }\label{eigen2}
\end{figure}

The effective Hamiltonian in the double-exciton subspace spanned by $\{|002\rangle, |101\rangle, |200\rangle\}$, where the magnon remains at the ground mode, can be written as
\begin{equation}\label{Heff2}
H_{\rm eff}^{(2)}=\begin{bmatrix}
0 & \sqrt{2}\tilde{G} & 0\\
\sqrt{2}\tilde{G} & 0 & \sqrt{2}\tilde{G}\\
0 & \sqrt{2}\tilde{G} & 0
\end{bmatrix},
\end{equation}
whose eigenstructure is found to be
\begin{equation}
\left\{
  \begin{array}{ll}
    E_1=\sqrt{2}\tilde{G}, & |\Psi_1\rangle=\frac{|002\rangle+|200\rangle}{2}+\frac{|101\rangle}{\sqrt{2}}, \\
    E_2=-\sqrt{2}\tilde{G}, & |\Psi_2\rangle=\frac{|002\rangle+|200\rangle}{2}-\frac{|101\rangle}{\sqrt{2}}, \\
    E_3=0, & |\Psi_3\rangle=\frac{|002\rangle-|200\rangle}{\sqrt{2}},
  \end{array}
\right.
\end{equation}
and the time-evolution operator reads
\begin{equation}
U(t)=\frac{1}{2}\begin{bmatrix}
c+1 & -s\sqrt{2}i & c-1\\
-s\sqrt{2}i & c & -s\sqrt{2}i\\
c-1 & -s\sqrt{2}i & c+1
\end{bmatrix}
\end{equation}
with $c\equiv\cos(2\tilde{G}t)$ and $s\equiv\sin(2\tilde{G}t)$. The energy diagram and the avoided level-crossing between $|200\rangle$ and $|002\rangle$ are plotted in Fig.~\ref{eigen2}. The conversion time is still $\pi/|2\tilde{G}|$. In addition, it is interesting to observe that the last eigenstate $|\Psi_3\rangle$ is a dark state in this specific case. The vanishing eigenvalue implies that this eigenstate will remain intact during the time evolution.

More versatile physics can be exploited in the subspace with more excitons, in which the magnon always stays as the vacuum state. For a fixed total number of photon and phonon $N$, the effective Hamiltonian in Eq.~(\ref{Heff}) in the subspace spanned by $\{|N00\rangle,|(N-1)01\rangle,\cdots,|10(N-1)\rangle,|00N\rangle\}$ can be expressed by
\begin{equation}\label{Heffn}
\begin{aligned}
H^{(N)}_{\rm eff}&=\tilde{G}\begin{bmatrix}
0 & L_1 & 0 & \cdot & 0\\
L_1 & 0 &L_2 & \cdot & 0\\
0 & L_2 & 0 & \cdot & 0\\
\cdot & \cdot & \cdot &\cdot & \cdot   \\
\cdot & \cdot & \cdot &\cdot &L_{N-1}\\
0 &0 & 0 & L_{N-1} & 0
\end{bmatrix},
\end{aligned}
\end{equation}
where $L_{n}=\sqrt{n(N-n)}$. It is interesting to be identical to the matrix representation of the Hamiltonian $H=2\tilde{G}S_x$, where $S_x$ is the angular momentum operator for a fictitious particle with spin $S=\frac{1}{2}(N-1)$. The Hamiltonian $H^{(N)}_{\rm eff}$ also describes an open-end spin chain governed by the site-dependent nearest-neighbor interactions~\cite{spin}
\begin{equation}\label{Hspin}
H_{\rm spin}=\sum_{n}\frac{\tilde{G}L_n}{2}(\sigma^x_n\sigma^x_{n+1}+\sigma^y_n\sigma^y_{n+1}),
\end{equation}
which is used to realize a perfect state transfer through the chain. If the initial state is prepared as $|N00\rangle$, then the probability amplitude for state transfer is
\begin{equation}\label{Ft}
A(t)=\langle00N|\exp(-i2\tilde{G}S_xt)|N00\rangle=[-i\sin(\tilde{G}t)]^{N-1}.
\end{equation}
Therefore, the perfect conversion about a quantum state between photon and phonon is accomplished in a constant time $\pi/|2\tilde{G}|$, irrespective to the exciton number or the particular subspace. Ideally, any superposed state of the photon can be converted into the phonon mode through the evolution time $\pi/|2\tilde{G}|$.

\section{The application range of the effective Hamiltonian}\label{fidelity}

\begin{figure}[htbp]
\centering
\includegraphics[width=0.4\textwidth]{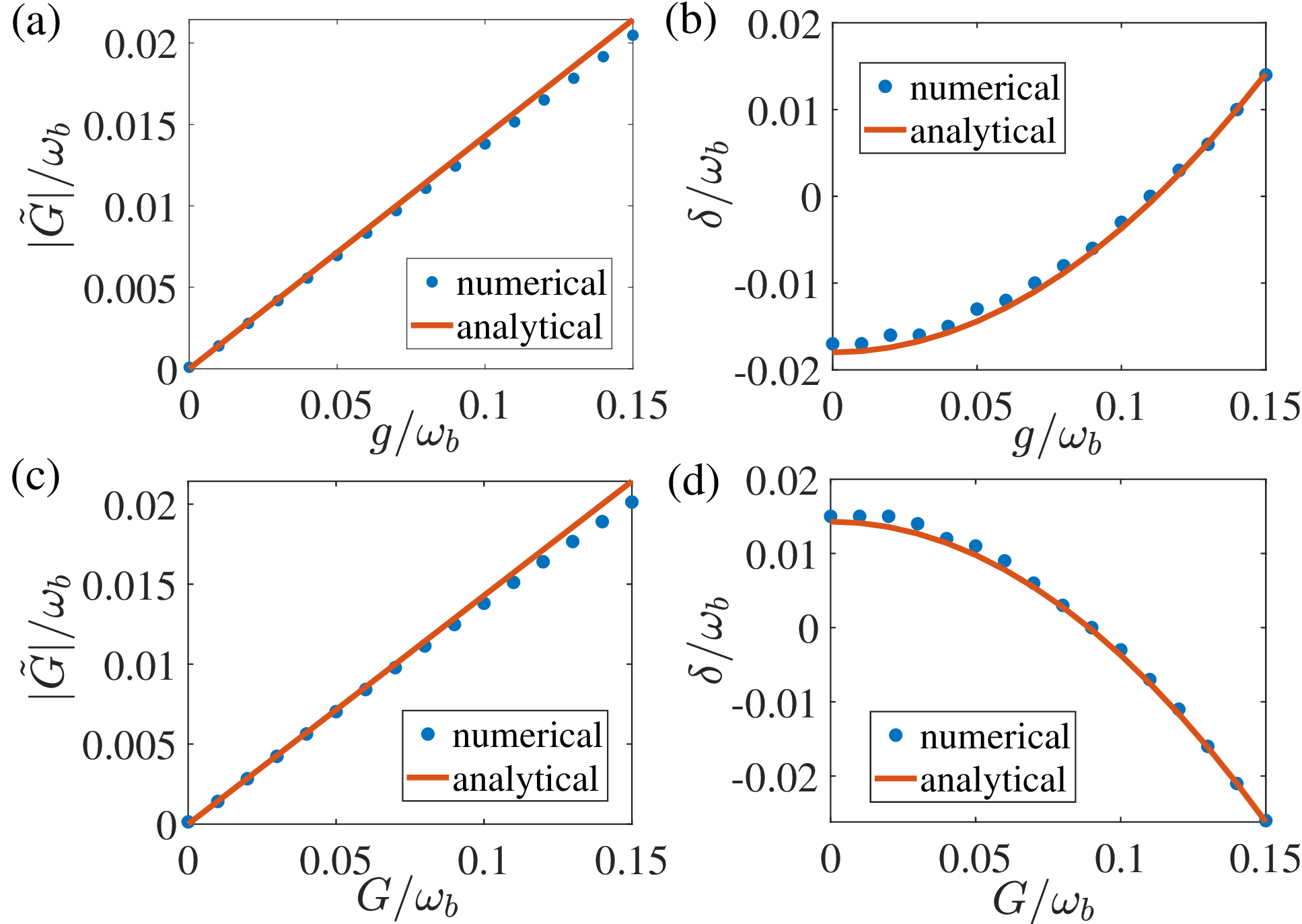}
\caption{(Color online) (a) and (c) Comparison between the numerically calculated normalized effective coupling strength $|\tilde{G}|/\omega_b$ (blue dots) and the corresponding analytical results in Eq.~(\ref{H}) from the second-order perturbation theory (orange solid lines) as functions of $g/\omega_b$ and $G/\omega_b$, respectively; (b) and (d) Comparison between the numerically calculated normalized energy shift $\delta/\omega_b$ (blue dots) and the corresponding analytical results in Eq.~(\ref{delta}) (orange solid lines) as functions of $g/\omega_b$ and $G/\omega_b$, respectively. For (a) and (b) $G=0.1\omega_b$, and for (c) and (d) $g=0.1\omega_b$. Here we fix $\Delta_m=1.7\omega_b$. }\label{gG1}
\end{figure}

We first check the applicability range of the effective Hamiltonian in Eq.~(\ref{Heff}) in terms of the coupling strength. It can be estimated or constrained by comparing the numerical results obtained from the original Hamiltonian in Eq.~(\ref{H}) for the full Hilbert space and the analytical results obtained via the perturbative derivation for the reduced subspace. The energy-splitting $|\tilde{G}|$ of the two eigenstates at the avoided level-crossing point (see Fig.~\ref{eigen1} in the single-exciton case) is presented in Fig.~\ref{gG1}(a) and (c) as a function of the original coupling strengthes $g$ and $G$ in the interaction Hamiltonian, respectively. The result given by the analytical expression in Eq.~(\ref{G}) is compared to that evaluated by the numerical simulation over the whole Hilbert space. It is found that the effective coupling strength $|\tilde{G}|$ is valid until $g/\omega_b\simeq0.1$ and $G/\omega_b\simeq0.1$. For a larger $g$ or $G$, higher-order contributions have to be included to capture the whole effect from the interaction Hamiltonian on modifying the eigen-structure of the bare system. However, an apparent yet still small deviation can be observed when either $g$ or $G$ is enhanced to $0.15\omega_b$. The effective Hamiltonian in Eq.~(\ref{Heff}) could therefore be used to investigate the effects in the strong-coupling regime.

Similarly, the energy shift $\delta$ in Eq.~(\ref{delta}) can also be justified by Fig.~\ref{gG1}(b) and (d). We check the same range of $g$ and $G$ as in Fig.~\ref{gG1}(a) and (c). It is shown that the analytical results do match with the numerical ones at least when the normalized photon-magnon interaction strength $g/\omega_b\leq0.15$ or phonon-magnon coupling strength $G/\omega_b\leq0.15$.

Next we take the open-quantum-system framework to further test the validness of the effective Hamiltonian in Eq.~(\ref{Heff}) in terms of the state-evolution fidelity. Under the standard assumptions (Markovian approximation, factorization of the system-environment density matrix, structure-free environment at the vacuum state), one can arrive at the Lindblad master equation for the density matrix of the interested hybrid-system,
\begin{equation}\label{mas}
\dot{\rho}(t)=-i[H,\rho(t)]+\kappa_a\mathcal{L}[a]\rho(t)+\kappa_m\mathcal{L}[m]\rho(t)+\gamma_b\mathcal{L}[b]\rho(t).
\end{equation}
Here $H$ is the full Hamiltonian of the system given in Eq.~(\ref{H}), $\kappa_a$, $\kappa_m$ and $\gamma_b$ are the relaxation rates for cavity-mode, magnon and phonon, respectively, and the superoperator $\mathcal{L}[O]$, $O=a,m,b$, is defined as
\begin{equation}\label{L}
\mathcal{L}[O]\rho\equiv\frac{1}{2}(2O\rho O^\dag-O^\dag O\rho-\rho O^\dag O).
\end{equation}
While for the effective model constituted by the photon and the phonon in Eq.~(\ref{Heff}), the Lindblad equation in the reduced subspace is written as
\begin{equation}\label{maseff}
\dot{\rho}(t)=-i[H_{\rm eff},\rho(t)]+\kappa_a\mathcal{L}[a]\rho(t)+\gamma_b\mathcal{L}[b]\rho(t).
\end{equation}
To simplify the discussion but with no loss of generality, we assume the decay rates of the photon and the magnon are the same $\kappa_a=\kappa_m=\kappa$, and set the decay rate of phonon $\gamma_b=10^{-2}\kappa$ regarding that the decoherence rates of phonon are much smaller than that of cavity-mode or magnon~\cite{mppentang}.

\begin{figure}[htbp]
\centering
\includegraphics[width=0.4\textwidth]{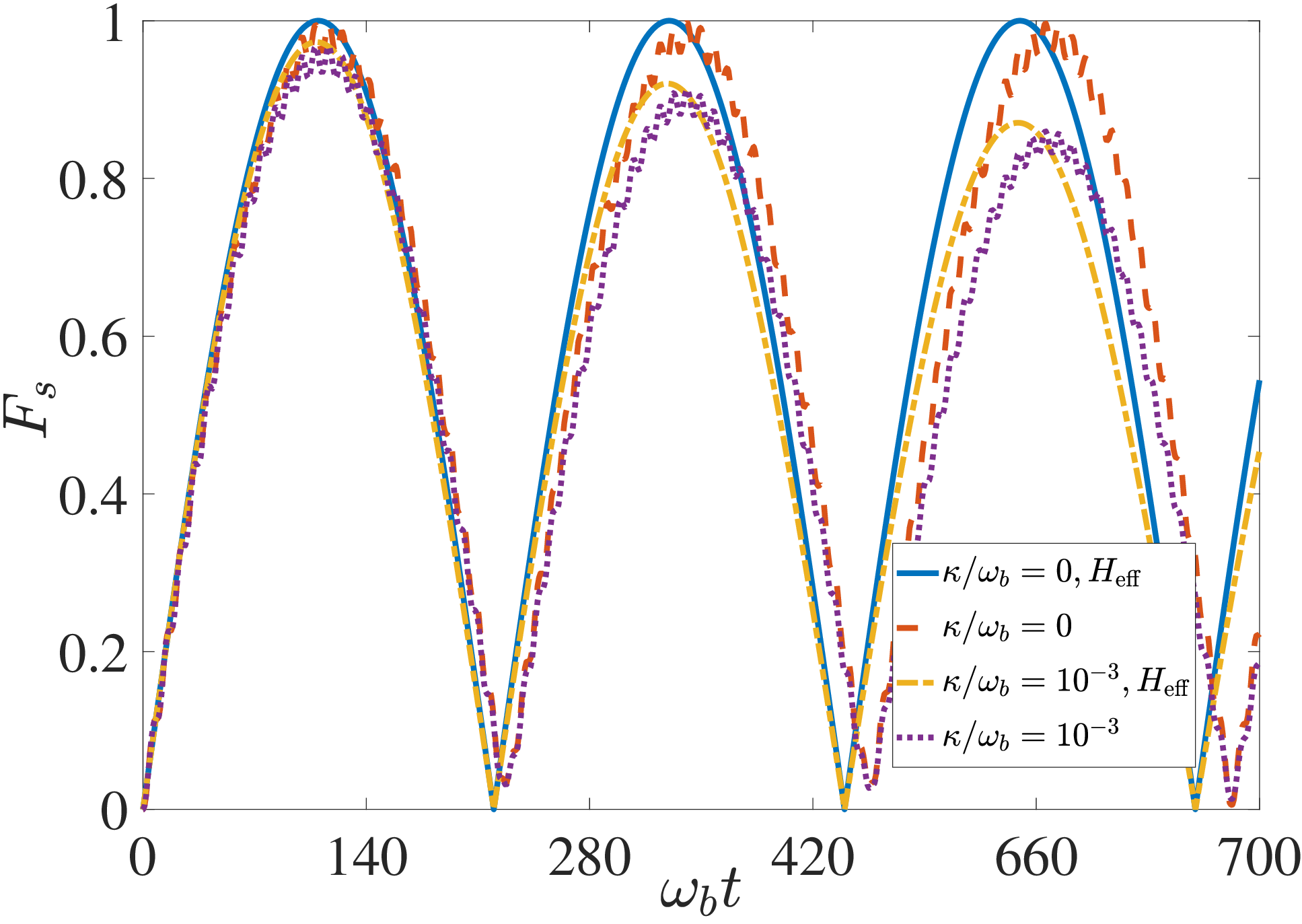}
\caption{(Color online) Time evolution of the state-fidelity by the master equations~(\ref{mas}) and (\ref{maseff}) to show the transfer between the initial state $|100\rangle$ and the target state $|001\rangle$. Here we fixed $\Delta_m=1.7\omega_b$ and $g=G=0.1\omega_b$.}\label{master}
\end{figure}

Then one can numerically calculate the state-fidelity $F_s=\sqrt{\langle\phi|\rho|\phi\rangle}$, where $|\phi\rangle$ is the target state, under either equation~(\ref{mas}) or equation~(\ref{maseff}). The distinction between the results from different master equation measures both the validness of the effective Hamiltonian and the robustness of our photon-phonon conversion protocol. In Fig.~\ref{master}, we plot the state-fidelity dynamics under different decoherence rates when the initial state and the target state are chosen as $|001\rangle$ and $|\phi\rangle=|100\rangle$, respectively. The blue solid line and the yellow dot-dashed line are the results from Eq.~\eqref{maseff} using the effective Hamiltonian, and the orange dashed line and the purple dotted line are those from Eq.~(\ref{mas}) using the full Hamiltonian. It is found that the results using the full Hamiltonian have a slightly longer period than those using the effective Hamiltonian in time evolution. Yet they match almost with each other at least for the first two or three periods of Rabi oscillation. In case of no decoherence, the state-fidelity achieves over $0.97$ by the full Hamiltonian. With the Markovian decoherence rate $\kappa=10^{-3}\omega_b$, both master equations produce the desired target-state with a fidelity over $0.90$ for photon-phonon conversion during the first period and maintain the target state with a fidelity about $0.85$ during the third period.

\section{The transfer fidelity under the structured environments}\label{dynamics}

The phenomenological analysis on the Markovian errors on the state-fidelity depends on the choice of the initial state and the target state during the photon-phonon conversion. It is indispensable to discuss the relationship between the transfer-fidelity by the effective Hamiltonian, that is independent on the state, and the non-Markovian errors, that is ubiquitous in almost all of the solid state systems. In this section, we investigate the effect from the structured environment using the non-Markovian Heisenberg-Langevin equation~\cite{optcavity,opttransfer}, by assuming that in a microscopic way the system is coupled to its environment consisted of a collection of independent harmonic oscillators. The total Hamiltonian reads,
\begin{equation}
H_{\rm tot}=H_{\rm eff}+H_E+H_I.
\end{equation}
Here the environmental Hamiltonian for the photon-phonon system reads
\begin{equation}\label{He}
H_E=\sum_k\omega_k a^\dag_k a_k+\sum_j\Omega_jb^\dag_jb_j,
\end{equation}
where $\omega_k$ and $\Omega_j$ are the reservoir frequency for the $k$th optical and the $j$th mechanical mode respectively. The two reservoirs are assumed to be uncorrelated. The interaction between the system and the environment can then be described by
\begin{equation}\label{Hi}
H_I=\sum_k g_k(a^\dag_k a+a^\dag a_k)+\sum_jf_j(b^\dag_jb+b^\dag b_j),
\end{equation}
where $g_k$ and $f_j$ are the respective system-reservoir coupling strength for the optical mode and the mechanical mode~\cite{nonmarkovmaster} and supposed to be real numbers for simplicity. Here the interaction Hamiltonian is written in a form under the rotating-wave approximation, which is valid when the coupling strength is much smaller than the resonant frequency of the system, i.e., $g_k,f_j\ll\omega_b,\Delta_a$.

With the effective Hamiltonian for the system part in Eq.~(\ref{Heff}) and the interaction Hamiltonian in Eq.~(\ref{Hi}), one can write down the Heisenberg-Langevin equations in the rotating frame with respect to the environmental Hamiltonian in Eq.~(\ref{He}),
\begin{equation}\label{O}
\dot{O}(t)=-iMO(t)-\int^t_0 d\tau \bar{F}(t-\tau) O(\tau) +\epsilon_{\rm in}(t).
\end{equation}
Here the time-evolution operator for the system modes is $O(t)=(a(t),b(t))^T$ and the input noise operator is $\epsilon_{\rm in}(t)=(a_{\rm in}(t),b_{\rm in}(t))^T$, where $a_{\rm in}\equiv-i\sum_kg_ke^{-i\omega_kt}a_k(0)$ and $b_{\rm in}\equiv-i\sum_jf_je^{-i\Omega_jt}b_j(0)$ depend on the initial condition of the environment. The thermal average occupation numbers in the particular modes of the two reservoirs are $\bar{n}_a(\omega_k)\equiv\langle a^\dag_k(0)a_{k'}(0)\rangle=1/[\exp(\hbar\omega_k/k_BT_a)-1]\delta_{kk'}$ and $\bar{n}_b(\omega_j)\equiv\langle b^\dag_j(0)b_{j'}(0)\rangle=1/[\exp(\hbar\omega_j/k_BT_b)-1]\delta_{jj'}$, respectively. The coefficient matrix $M$ and $\bar{F}(t)$ are given by
\begin{equation}\label{MF}
M=\begin{pmatrix}
0 & \tilde{G}\\
\tilde{G} & 0
\end{pmatrix}, \quad
\bar{F}(t)=\begin{pmatrix}
f_a(t)& 0\\
0 & f_b(t)
\end{pmatrix},
\end{equation}
respectively, where $f_a(t)\equiv\sum_kg^2_ke^{-i\omega_k t}$ and $f_b(t)\equiv\sum_jf_j^2e^{-i\Omega_jt}$ are the non-local time correlation functions. We can rewrite the correlation functions by introducing the spectral density functions
\begin{equation}
\label{fab}
f_x(t)=\int d\omega J_x(\omega)e^{-i\omega t}, \quad x=a,b
\end{equation}
for the optical and the mechanical environments, respectively. The environment for the cavity mode is generally of an Ohmic-like spectrum, i.e., $J_a(\omega)=\eta\omega(\omega/\omega_0)^{s-1}e^{-\omega/\omega_0}$, where $\eta$ is a dimensionless coupling strength between system and environment, and $\omega_0$ is a high-frequency cutoff~\cite{spectrum}. The parameter $s$ classifies the environment as sub-Ohmic ($0<s<1$), Ohmic ($s=1$), or super-Ohmic ($s>1$). For the phonon mode in the YIG sphere, the solid-state environment is assumed to be a $1/f$-like spectrum, similar to that for the optomechanical system recently measured in experiments. We therefore use the spectral density function $J_b(\omega)= C\omega^k$, where the coupling coefficient $C>0$, $k$ is a negative number around $-1$~\cite{spectrum2} and the bandwidth in this work is chosen as $\omega\in(0.1\omega_b, 2\omega_b)$.

Formally, Eq.~(\ref{O}) can be solved by assuming $O(t)=\mathcal{U}(t)O(0)+\mathcal{V}(t)$, where $\mathcal{U}(t)$ is a $2\times 2$ coefficient matrix $(\mathcal{U}_{11}(t), \mathcal{U}_{12}(t); \mathcal{U}_{21}(t), \mathcal{U}_{22}(t))$ as a function of time, and $\mathcal{V}(t)$ is a vector of operators $(\mathcal{V}_1(t), \mathcal{V}_2(t))^T$ related to the nonequilibrium Green's functions of the system. These Green's functions obey the following Dyson equations
\begin{equation}\label{uv}
\begin{aligned}
\dot{\mathcal{U}}(t)=-iM\mathcal{U}(t)-\int^t_0d\tau \bar{F}(t-\tau)\mathcal{U}(\tau), \\
\dot{\mathcal{V}}(t)=-iM\mathcal{V}(t)-\int^t_0d\tau \bar{F}(t-\tau)\mathcal{V}(\tau)+\epsilon_{\rm in}(t).
\end{aligned}
\end{equation}
Considering the initial conditions $\mathcal{U}(0)=\mathcal{I}$ and $\mathcal{V}(0)=0$, one can formally have
\begin{equation}\label{V}
\mathcal{V}(t)=\int^t_0d\tau\mathcal{U}(t-\tau)\epsilon_{\rm in}(\tau).
\end{equation}
With the solution about $\mathcal{U}(t)$, the dynamical evolution of $b(t)$ is written as
\begin{equation}\label{at}
b(t)=\mathcal{U}_{21}(t)a(0)+\mathcal{U}_{22}(t)b(0)+\mathcal{V}_{2}(t).
\end{equation}
Then the expectation value of the phonon number $\langle b^\dag(t)b(t)\rangle$ can be evaluated by
\begin{eqnarray}\nonumber
\langle b^\dag(t)b(t)\rangle&=&|\mathcal{U}_{21}(t)|^2\langle a^\dag(0)a(0)\rangle+|\mathcal{U}_{22}(t)|^2\langle b^\dag(0)b(0)\rangle\\ \label{Nb} &+& \langle\mathcal{V}^\dag_2(t)\mathcal{V}_2(t)\rangle.
\end{eqnarray}
The first term on the right-hand side of Eq.~(\ref{Nb}) is proportional to the initial average number of photon $\langle a^\dag(0)a(0)\rangle$. The second term is proportional to the initial average number of phonon $\langle b^\dag(0)b(0)\rangle$, whose contribution could be reduced by pre-cooling the mechanical oscillator to the ground state. As for the last term, one can express it by
\begin{equation}\label{v2}
\begin{aligned}
\langle \mathcal{V}^\dag_2 \mathcal{V}_2\rangle&=\int \frac{d\omega}{2\pi} J_a(\omega) \bar{n}_a (\omega)\left|\int^t_0 \mathcal{U}_{21} (t-\tau) e^{-i\omega\tau}\right|^2\\
&+\int \frac{d\omega}{2\pi} J_b(\omega) \bar{n}_b (\omega)\left|\int^t_0 \mathcal{U}_{22} (t-\tau) e^{-i\omega\tau}\right|^2.
\end{aligned}
\end{equation}
Assume the environmental temperature for the photon mode $T_a\approx0$ and the bandwidth of the spectral density for the mechanical environment is sufficiently narrow around the resonant frequency, the average value $\langle\mathcal{V}^\dag_2\mathcal{V}_2\rangle$ is upper-bounded by $\bar{n}_b(\omega_b)[\mathcal{V}_2, \mathcal{V}^\dag_2]$ due to the fact that
\begin{equation}\label{V2}
\begin{aligned}
\left[\mathcal{V}_2, \mathcal{V}^\dag_2\right]&=\int \frac{d\omega}{2\pi}J_a(\omega)\left|\int^t_0 \mathcal{U}_{21}(t-\tau) e^{-i\omega \tau}\right|^2\\ &+\int \frac{d\omega}{2\pi}J_b(\omega)\left|\int^t_0\mathcal{U}_{22}(t-\tau) e^{-i\omega \tau}\right|^2.
\end{aligned}
\end{equation}
Also note the commutation relation $[b(t), b^\dag(t)]=1$ holds for any moment $t$, which renders
\begin{equation}\label{uv2}
|\mathcal{U}_{21}|^2+|\mathcal{U}_{22}|^2+[\mathcal{V}_2, \mathcal{V}^\dag_2]=1.
\end{equation}
Apparently $[\mathcal{V}_2, \mathcal{V}^\dag_2]\ge0$. It means that under a low temperature $T_b$, both $|\mathcal{U}_{22}|$ and $\langle\mathcal{V}^\dag_2\mathcal{V}_2\rangle$ are close to zero when $|\mathcal{U}_{21}|\approx1$. We therefore understand that the transfer fidelity $F$ from photon to phonon can be quantified by $|\mathcal{U}_{21}(t)|$~\cite{opttransfer} using the numerical solution of Eq.~(\ref{uv}). Since this fidelity is obtained in the Heisenberg picture, it is then independent of the initial and target states.

\begin{figure}[htbp]
\centering
\includegraphics[width=0.4\textwidth]{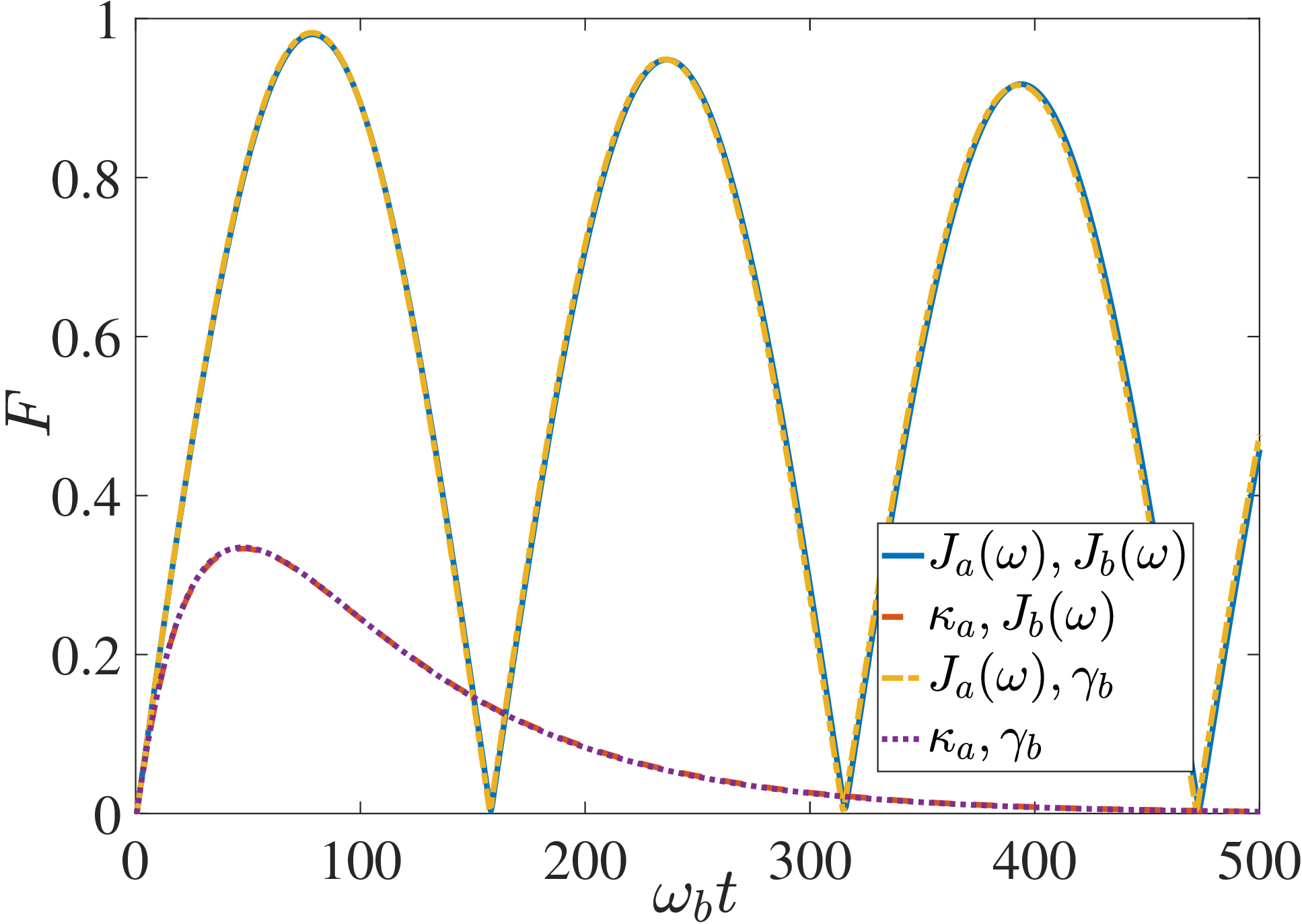}
\caption{(Color online) The time evolution of the transfer fidelity under various environments. The parameters for the structured environments are chosen as $|\tilde{G}|=0.02\omega_b$, $\eta=1\times10^{-4}\omega_b$, $\omega_0=5\omega_b$, $s=1$, $C=1\times 10^{-4}\omega_b$ and $k=-1$. For the Markovian environments, the decay rates $\kappa_a$ and $\gamma_b$ are obtained by the Weisskopf-Wigner theory. }\label{mar}
\end{figure}

We first comparing the noise effects on $F$ from various combinations of the structured and the Markovian environments for the photon and phonon modes as shown in Fig.~\ref{mar}. The structured environments are characterized by their spectral functions and the Markovian environments are characterized by their decay constants, which are obtained under the Weisskopf-Wigner approximation. In particularly, the optical-mode decay rate is $\kappa_a=J_a(\omega_a)/2$ and the mechanical-mode decay rate is $\gamma_b=J_b(\omega_b)/2$. When both modes are embedded in the realistic structured environments (see the blue-solid line), the conversion fidelity is close to $0.98$ in the first period of Rabi oscillation and could be still maintained above $0.9$ in the third period. It is nearly invariant when the environment for the phonon mode is changed to be a Markovian type (see the yellow-dotted line). The conversion will be greatly damaged when the photon mode is subject to a Markovian type and in this case, it is also insensitive to the choice of the phonon-mode environment (see the red-dashed the purple-dotted lines). In both cases, the conversion fidelity is no more than $0.35$ and could not start a second period of Rabi oscillation. These results indicate that the non-Markovian noise from the structured environment for the photon mode are more significant than that for the phonon mode on protecting the conversion fidelity.

\begin{figure}[htbp]
\centering
\includegraphics[width=0.4\textwidth]{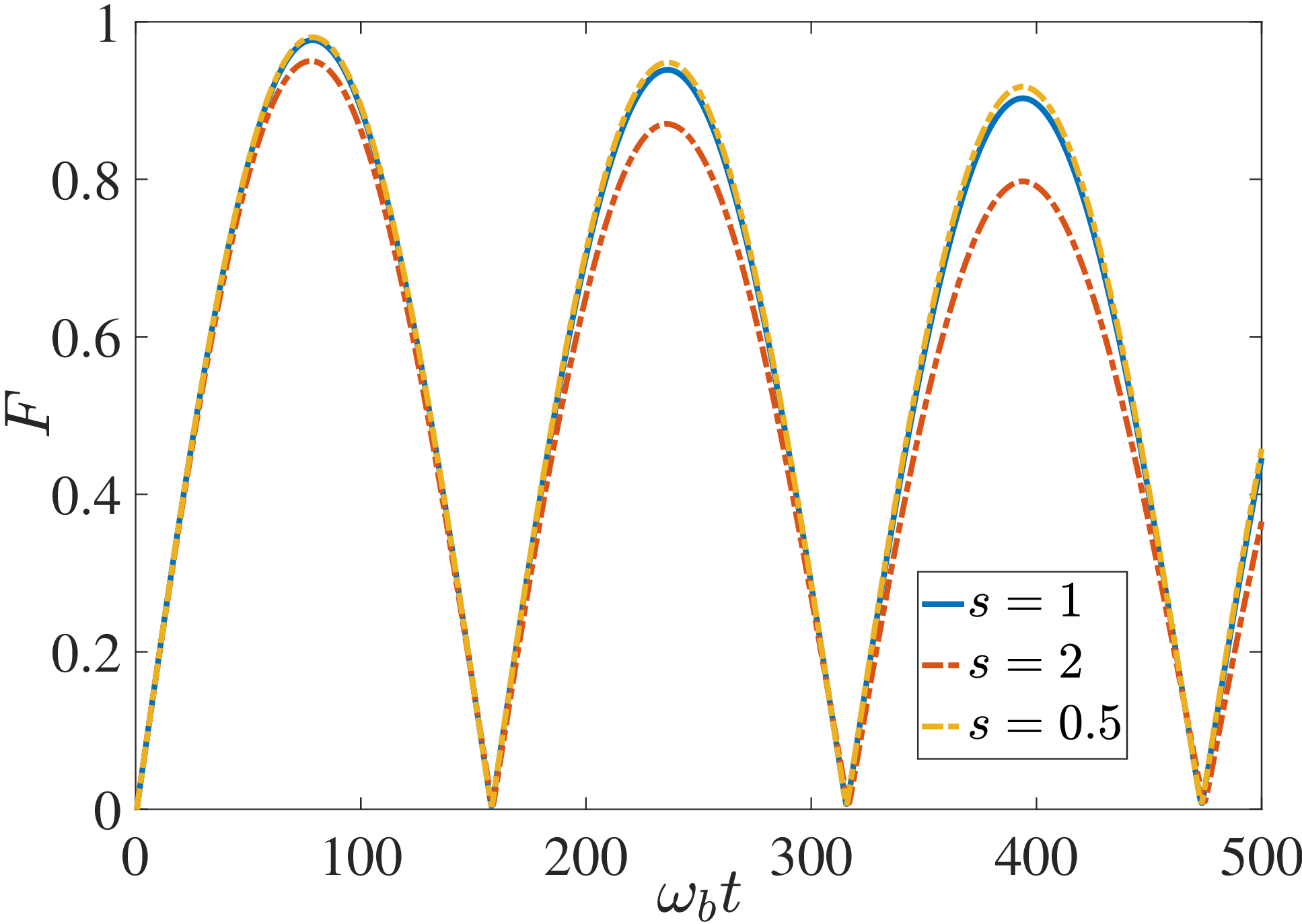}
\caption{(Color online) The time evolution of the transfer fidelity under sub-Ohmic ($s<1$), Ohmic ($s=1$), and super-Ohmic ($s>1$) environments for the photon mode and a fixed $1/f$-noise environment for the phonon mode. The other parameters are set as $|\tilde{G}|=0.02\omega_b$, $\eta=1\times 10^{-4}\omega_b$, $\omega_0=5\omega_b$, and $C=1\times 10^{-4}\omega_b$.}\label{state}
\end{figure}

In Fig.~\ref{state}, the phonon mode is subjected to a fixed $1/f$ noise with $k=-1$ and then one can observe the effects from the sub-Ohmic ($s=0.5$), Ohmic ($s=1$), and super-Ohmic ($s=2$) environments for the photon mode on the time evolution of the transfer fidelity. The other parameters of these power-law noises are fixed. We can see that the sub-Ohmic spectrum and the Ohmic spectrum are better than the super-Ohmic spectrum in terms of conversion fidelity. For the sub-Ohmic spectrum, the fidelity could be close to $0.99$ during the first period of Rabi oscillation and still around $0.90$ during the third period. These results are close to those for the Ohmic spectrum. While the fidelity will drop to about $0.80$ during the third period of Rabi oscillation for the super-Ohmic spectrum.

\begin{figure}[htbp]
\centering
\includegraphics[width=0.4\textwidth]{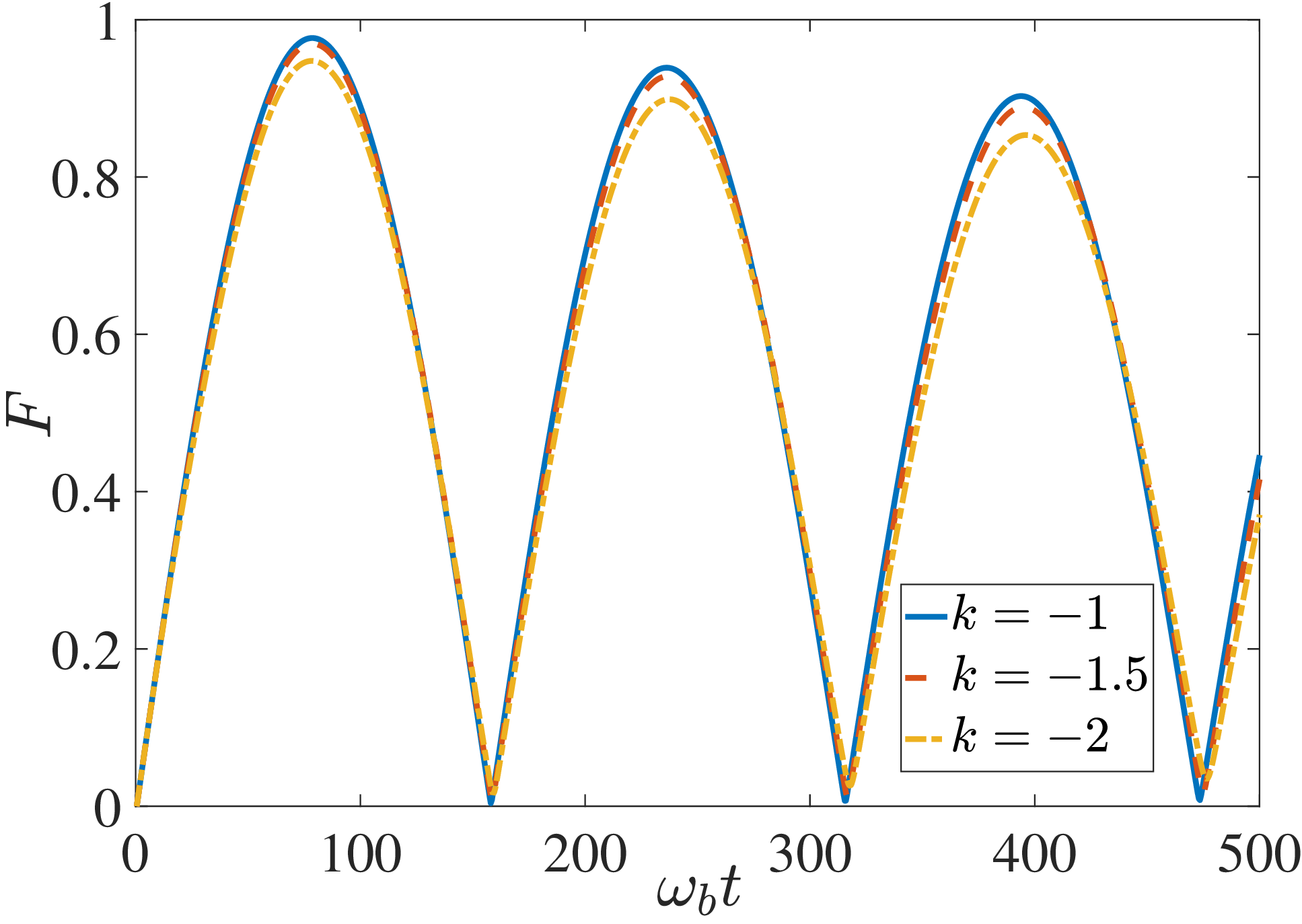}
\caption{(Color online) The time evolution of the transfer fidelity under various environments for the phonon mode and a fixed structured environment for the photon mode with an Ohmic spectrum. The other parameters are set as $|\tilde{G}|=0.02\omega_b$, $\eta=1\times 10^{-4}\omega_b$, $\omega_0=5\omega_b$, and $C=1\times 10^{-4}\omega_b$.}\label{ck}
\end{figure}

In Fig.~\ref{ck}, one can observe that the conversion fidelity is also insensitive to different choice of the structured environments for the phonon mode. Under a fixed Ohmic environment for the photon mode, the conversion fidelity shows a slightly decreasing pattern with increasing $|k|$. Thus in realistic situations, the conversion process between the photon mode and the phonon mode is strongly resilient to the $1/f$-type noise for the phonon mode.

\section{Discussion and conclusion}\label{diss}

When the yttrium iron garnet sphere are strongly pumped to generate a considerable number of magnons, the Kerr effect due to the ensued magnetocrystalline anisotropy could not be completely neglected. Considering that the exciton number in the spin wave is much smaller than the total number of spins in YIG sphere, one can modify the system Hamiltonian in Eq.~(\ref{Hs}) to be~\cite{kerr1,kerr2},
\begin{equation}\label{Hss}
\tilde{H}_S=H_S+Km^\dag m m^\dag m,
\end{equation}
where $K$ is the nonlinear coefficient for the Kerr effect. With the linear approximation, the Kerr term $Km^\dag m m^\dag m$ becomes $2K\langle m^\dag m\rangle m^\dag m$, thus the Hamiltonian $H_0$ in Eq.~(\ref{H}) turns out to be
\begin{equation}\label{H0s}
\tilde{H}_0=\Delta_aa^{\dag}a+\tilde{\Delta}_mm^{\dag}m+\omega_b b^{\dag}b,
\end{equation}
where $\tilde{\Delta}_m\equiv\Delta_m+2K\langle m^\dag m\rangle$~\cite{kerr3}. Then the derivation process for the effective Hamiltonian describing the photon-phonon conversion [from Eq.~(\ref{H}) to Eq.~(\ref{Heff})] would be merely modified in quantity to ensure that the frequency detuning of the magnon-mode approaches the frequency of the cavity-mode. The Kerr term will eventually only change the effective coupling strength $\tilde{G}$ in Eq.~(\ref{G}). For a YIG sphere with a diameter about $250$ $\mu$m, the Kerr coefficient $K/2\pi\approx6.4\times10^{-9}$ Hz and the magnon exciton number $\langle m^\dag m\rangle\approx10^{14}$ corresponding to $G=Mg_{mb}\sim10^7$ Hz via $M=\sqrt{\langle m^\dag m\rangle}$. Then one can estimate that $K\langle m^\dag m\rangle\approx 10^5$ Hz, which is much smaller than $\Delta_m$. It is therefore believed that our proposal also adapts to such a hybrid system even with a non-negligible Kerr effect.

In summary, we have presented a protocol to realize a photon-phonon interface or conversion in a cavity magnomechanical system, where the magnon mode in the YIG sphere is coupled to both a microwave cavity mode and the mechanical vibration mode in the same sphere. Our indirect photon-phonon-conversion protocol has considerable merit in physics due to the strong controllability in this novel system. This magnon-assisted protocol relies on the effective Hamiltonian for coupling photon and phonon, which is constructed in the strong coupling regimes for both magnon-photon interaction and magnon-phonon interaction. We apply a bottom-up perturbation approach rather than the adiabatic-elimination technique, which is beneficial to shed light on more physics in special subspaces. In the open-quantum-system framework, the state fidelity of the photon-phonon conversion is mainly limited by the decay of the microwave cavity $\kappa_a$ rather than the dissipation of the phonons $\gamma_b$, due to the fact that $\gamma_b$ is much smaller than the phonon frequency $\omega_b$~\cite{mppentang}. We also analyze the effect from various environmental noises on the conversion fidelity using the non-Markovian Heisenberg-Langevin equations. Excellent results can be obtained when the cavity mode is under a structured environment with sub-Ohmic spectrum even if the phonons are in a Markovian environment. Our work in pursuit of the quantum state transfer and protection therefore provides a novel implementation of the photon-phonon interface realized in a solid system under realistic noises. It also extends the application range of the cavity-magnomechanical system as a promising hybrid platform for quantum information processing.

\section*{Acknowledgments}

We acknowledge grant support from the National Science Foundation of China (Grants No. 11974311 and No. U1801661), Zhejiang Provincial Natural Science Foundation of China under Grant No. LD18A040001, and the Fundamental Research Funds for the Central Universities (No. 2018QNA3004).

\bibliographystyle{apsrevlong}
\bibliography{reference}

\begin{thebibliography}{51}%
\makeatletter
\providecommand \@ifxundefined [1]{%
 \@ifx{#1\undefined}
}%
\providecommand \@ifnum [1]{%
 \ifnum #1\expandafter \@firstoftwo
 \else \expandafter \@secondoftwo
 \fi
}%
\providecommand \@ifx [1]{%
 \ifx #1\expandafter \@firstoftwo
 \else \expandafter \@secondoftwo
 \fi
}%
\providecommand \natexlab [1]{#1}%
\providecommand \enquote  [1]{``#1''}%
\providecommand \bibnamefont  [1]{#1}%
\providecommand \bibfnamefont [1]{#1}%
\providecommand \citenamefont [1]{#1}%
\providecommand \href@noop [0]{\@secondoftwo}%
\providecommand \href [0]{\begingroup \@sanitize@url \@href}%
\providecommand \@href[1]{\@@startlink{#1}\@@href}%
\providecommand \@@href[1]{\endgroup#1\@@endlink}%
\providecommand \@sanitize@url [0]{\catcode `\\12\catcode `\$12\catcode
  `\&12\catcode `\#12\catcode `\^12\catcode `\_12\catcode `\%12\relax}%
\providecommand \@@startlink[1]{}%
\providecommand \@@endlink[0]{}%
\providecommand \url  [0]{\begingroup\@sanitize@url \@url }%
\providecommand \@url [1]{\endgroup\@href {#1}{\urlprefix }}%
\providecommand \urlprefix  [0]{URL }%
\providecommand \Eprint [0]{\href }%
\providecommand \doibase [0]{http://dx.doi.org/}%
\providecommand \selectlanguage [0]{\@gobble}%
\providecommand \bibinfo  [0]{\@secondoftwo}%
\providecommand \bibfield  [0]{\@secondoftwo}%
\providecommand \translation [1]{[#1]}%
\providecommand \BibitemOpen [0]{}%
\providecommand \bibitemStop [0]{}%
\providecommand \bibitemNoStop [0]{.\EOS\space}%
\providecommand \EOS [0]{\spacefactor3000\relax}%
\providecommand \BibitemShut  [1]{\csname bibitem#1\endcsname}%
\let\auto@bib@innerbib\@empty
\bibitem [{\citenamefont {Lachance-Quirion}\ \emph {et~al.}(2019)\citenamefont
  {Lachance-Quirion}, \citenamefont {Tabuchi}, \citenamefont {Gloppe},
  \citenamefont {Usami},\ and\ \citenamefont {Nakamura}}]{magnon}%
  \BibitemOpen
  \bibfield  {author} {\bibinfo {author} {\bibfnamefont {D.}~\bibnamefont
  {Lachance-Quirion}}, \bibinfo {author} {\bibfnamefont {Y.}~\bibnamefont
  {Tabuchi}}, \bibinfo {author} {\bibfnamefont {A.}~\bibnamefont {Gloppe}},
  \bibinfo {author} {\bibfnamefont {K.}~\bibnamefont {Usami}}, \ and\ \bibinfo
  {author} {\bibfnamefont {Y.}~\bibnamefont {Nakamura}},\ }\bibfield  {title}
  {\emph {\bibinfo {title} {Hybrid quantum systems based on magnonics},\
  }}\href {\doibase 10.7567/1882-0786/ab248d} {\bibfield  {journal} {\bibinfo
  {journal} {Applied Physics Express}\ }\textbf {\bibinfo {volume} {12}},\
  \bibinfo {pages} {070101} (\bibinfo {year} {2019})}\BibitemShut {NoStop}%
\bibitem [{\citenamefont {Elyasi}\ \emph {et~al.}(2020)\citenamefont {Elyasi},
  \citenamefont {Blanter},\ and\ \citenamefont {Bauer}}]{magnon2}%
  \BibitemOpen
  \bibfield  {author} {\bibinfo {author} {\bibfnamefont {M.}~\bibnamefont
  {Elyasi}}, \bibinfo {author} {\bibfnamefont {Y.~M.}\ \bibnamefont {Blanter}},
  \ and\ \bibinfo {author} {\bibfnamefont {G.~E.~W.}\ \bibnamefont {Bauer}},\
  }\bibfield  {title} {\emph {\bibinfo {title} {Resources of nonlinear cavity
  magnonics for quantum information},\ }}\href {\doibase
  10.1103/PhysRevB.101.054402} {\bibfield  {journal} {\bibinfo  {journal}
  {Phys. Rev. B}\ }\textbf {\bibinfo {volume} {101}},\ \bibinfo {pages}
  {054402} (\bibinfo {year} {2020})}\BibitemShut {NoStop}%
\bibitem [{\citenamefont {Li}\ \emph {et~al.}(2020)\citenamefont {Li},
  \citenamefont {Zhang}, \citenamefont {Tyberkevych}, \citenamefont {Kwok},\
  and\ \citenamefont {Novosad}}]{magnon3}%
  \BibitemOpen
  \bibfield  {author} {\bibinfo {author} {\bibfnamefont {Y.}~\bibnamefont
  {Li}}, \bibinfo {author} {\bibfnamefont {W.}~\bibnamefont {Zhang}}, \bibinfo
  {author} {\bibfnamefont {V.}~\bibnamefont {Tyberkevych}}, \bibinfo {author}
  {\bibfnamefont {W.~K.}\ \bibnamefont {Kwok}}, \ and\ \bibinfo {author}
  {\bibfnamefont {V.}~\bibnamefont {Novosad}},\ }\bibfield  {title} {\emph
  {\bibinfo {title} {Hybrid magnonics: physics, circuits, and applications for
  coherent information processing},\ }}\href@noop {} {\bibfield  {journal}
  {\bibinfo  {journal} {J. Appl. Phys.}\ }\textbf {\bibinfo {volume} {128}},\
  \bibinfo {pages} {130902} (\bibinfo {year} {2020})}\BibitemShut {NoStop}%
\bibitem [{\citenamefont {Potts}\ and\ \citenamefont {Davis}(2020)}]{magnon4}%
  \BibitemOpen
  \bibfield  {author} {\bibinfo {author} {\bibfnamefont {C.~A.}\ \bibnamefont
  {Potts}}\ and\ \bibinfo {author} {\bibfnamefont {J.~P.}\ \bibnamefont
  {Davis}},\ }\bibfield  {title} {\emph {\bibinfo {title} {Strong magnon-photon
  coupling within a tunable cryogenic microwave cavity},\ }}\href@noop {}
  {\bibfield  {journal} {\bibinfo  {journal} {Appl. Phys. Lett.}\ }\textbf
  {\bibinfo {volume} {116}},\ \bibinfo {pages} {263503} (\bibinfo {year}
  {2020})}\BibitemShut {NoStop}%
\bibitem [{\citenamefont {Soykal}\ and\ \citenamefont
  {Flatt\'e}(2010{\natexlab{a}})}]{yigcavity1}%
  \BibitemOpen
  \bibfield  {author} {\bibinfo {author} {\bibfnamefont {O.~O.}\ \bibnamefont
  {Soykal}}\ and\ \bibinfo {author} {\bibfnamefont {M.~E.}\ \bibnamefont
  {Flatt\'e}},\ }\bibfield  {title} {\emph {\bibinfo {title} {Strong field
  interactions between a nanomagnet and a photonic cavity},\ }}\href {\doibase
  10.1103/PhysRevLett.104.077202} {\bibfield  {journal} {\bibinfo  {journal}
  {Phys. Rev. Lett.}\ }\textbf {\bibinfo {volume} {104}},\ \bibinfo {pages}
  {077202} (\bibinfo {year} {2010}{\natexlab{a}})}\BibitemShut {NoStop}%
\bibitem [{\citenamefont {Soykal}\ and\ \citenamefont
  {Flatt\'e}(2010{\natexlab{b}})}]{yigcavity2}%
  \BibitemOpen
  \bibfield  {author} {\bibinfo {author} {\bibfnamefont {O.~O.}\ \bibnamefont
  {Soykal}}\ and\ \bibinfo {author} {\bibfnamefont {M.~E.}\ \bibnamefont
  {Flatt\'e}},\ }\bibfield  {title} {\emph {\bibinfo {title} {Size dependence
  of strong coupling between nanomagnets and photonic cavities},\ }}\href
  {\doibase 10.1103/PhysRevB.82.104413} {\bibfield  {journal} {\bibinfo
  {journal} {Phys. Rev. B}\ }\textbf {\bibinfo {volume} {82}},\ \bibinfo
  {pages} {104413} (\bibinfo {year} {2010}{\natexlab{b}})}\BibitemShut
  {NoStop}%
\bibitem [{\citenamefont {Yuan}\ \emph {et~al.}(2020)\citenamefont {Yuan},
  \citenamefont {Yan}, \citenamefont {Zheng}, \citenamefont {He}, \citenamefont
  {Xia},\ and\ \citenamefont {Yung}}]{yigcavity3}%
  \BibitemOpen
  \bibfield  {author} {\bibinfo {author} {\bibfnamefont {H.~Y.}\ \bibnamefont
  {Yuan}}, \bibinfo {author} {\bibfnamefont {P.}~\bibnamefont {Yan}}, \bibinfo
  {author} {\bibfnamefont {S.}~\bibnamefont {Zheng}}, \bibinfo {author}
  {\bibfnamefont {Q.~Y.}\ \bibnamefont {He}}, \bibinfo {author} {\bibfnamefont
  {K.}~\bibnamefont {Xia}}, \ and\ \bibinfo {author} {\bibfnamefont {M.-H.}\
  \bibnamefont {Yung}},\ }\bibfield  {title} {\emph {\bibinfo {title} {Steady
  bell state generation via magnon-photon coupling},\ }}\href {\doibase
  10.1103/PhysRevLett.124.053602} {\bibfield  {journal} {\bibinfo  {journal}
  {Phys. Rev. Lett.}\ }\textbf {\bibinfo {volume} {124}},\ \bibinfo {pages}
  {053602} (\bibinfo {year} {2020})}\BibitemShut {NoStop}%
\bibitem [{\citenamefont {Liu}\ \emph {et~al.}(2019)\citenamefont {Liu},
  \citenamefont {Xiong},\ and\ \citenamefont {Wu}}]{magnonblockade}%
  \BibitemOpen
  \bibfield  {author} {\bibinfo {author} {\bibfnamefont {Z.-X.}\ \bibnamefont
  {Liu}}, \bibinfo {author} {\bibfnamefont {H.}~\bibnamefont {Xiong}}, \ and\
  \bibinfo {author} {\bibfnamefont {Y.}~\bibnamefont {Wu}},\ }\bibfield
  {title} {\emph {\bibinfo {title} {Magnon blockade in a hybrid
  ferromagnet-superconductor quantum system},\ }}\href {\doibase
  10.1103/PhysRevB.100.134421} {\bibfield  {journal} {\bibinfo  {journal}
  {Phys. Rev. B}\ }\textbf {\bibinfo {volume} {100}},\ \bibinfo {pages}
  {134421} (\bibinfo {year} {2019})}\BibitemShut {NoStop}%
\bibitem [{\citenamefont {Tabuchi}\ \emph {et~al.}(2014)\citenamefont
  {Tabuchi}, \citenamefont {Ishino}, \citenamefont {Ishikawa}, \citenamefont
  {Yamazaki}, \citenamefont {Usami},\ and\ \citenamefont
  {Nakamura}}]{yigcavity4}%
  \BibitemOpen
  \bibfield  {author} {\bibinfo {author} {\bibfnamefont {Y.}~\bibnamefont
  {Tabuchi}}, \bibinfo {author} {\bibfnamefont {S.}~\bibnamefont {Ishino}},
  \bibinfo {author} {\bibfnamefont {T.}~\bibnamefont {Ishikawa}}, \bibinfo
  {author} {\bibfnamefont {R.}~\bibnamefont {Yamazaki}}, \bibinfo {author}
  {\bibfnamefont {K.}~\bibnamefont {Usami}}, \ and\ \bibinfo {author}
  {\bibfnamefont {Y.}~\bibnamefont {Nakamura}},\ }\bibfield  {title} {\emph
  {\bibinfo {title} {Hybridizing ferromagnetic magnons and microwave photons in
  the quantum limit},\ }}\href {\doibase 10.1103/PhysRevLett.113.083603}
  {\bibfield  {journal} {\bibinfo  {journal} {Phys. Rev. Lett.}\ }\textbf
  {\bibinfo {volume} {113}},\ \bibinfo {pages} {083603} (\bibinfo {year}
  {2014})}\BibitemShut {NoStop}%
\bibitem [{\citenamefont {Zhang}\ \emph {et~al.}(2014)\citenamefont {Zhang},
  \citenamefont {Zou}, \citenamefont {Jiang},\ and\ \citenamefont
  {Tang}}]{yigcavity5}%
  \BibitemOpen
  \bibfield  {author} {\bibinfo {author} {\bibfnamefont {X.}~\bibnamefont
  {Zhang}}, \bibinfo {author} {\bibfnamefont {C.-L.}\ \bibnamefont {Zou}},
  \bibinfo {author} {\bibfnamefont {L.}~\bibnamefont {Jiang}}, \ and\ \bibinfo
  {author} {\bibfnamefont {H.~X.}\ \bibnamefont {Tang}},\ }\bibfield  {title}
  {\emph {\bibinfo {title} {Strongly coupled magnons and cavity microwave
  photons},\ }}\href {\doibase 10.1103/PhysRevLett.113.156401} {\bibfield
  {journal} {\bibinfo  {journal} {Phys. Rev. Lett.}\ }\textbf {\bibinfo
  {volume} {113}},\ \bibinfo {pages} {156401} (\bibinfo {year}
  {2014})}\BibitemShut {NoStop}%
\bibitem [{\citenamefont {Goryachev}\ \emph {et~al.}(2014)\citenamefont
  {Goryachev}, \citenamefont {Farr}, \citenamefont {Creedon}, \citenamefont
  {Fan}, \citenamefont {Kostylev},\ and\ \citenamefont {Tobar}}]{yigcavity6}%
  \BibitemOpen
  \bibfield  {author} {\bibinfo {author} {\bibfnamefont {M.}~\bibnamefont
  {Goryachev}}, \bibinfo {author} {\bibfnamefont {W.~G.}\ \bibnamefont {Farr}},
  \bibinfo {author} {\bibfnamefont {D.~L.}\ \bibnamefont {Creedon}}, \bibinfo
  {author} {\bibfnamefont {Y.}~\bibnamefont {Fan}}, \bibinfo {author}
  {\bibfnamefont {M.}~\bibnamefont {Kostylev}}, \ and\ \bibinfo {author}
  {\bibfnamefont {M.~E.}\ \bibnamefont {Tobar}},\ }\bibfield  {title} {\emph
  {\bibinfo {title} {High-cooperativity cavity qed with magnons at microwave
  frequencies},\ }}\href {\doibase 10.1103/PhysRevApplied.2.054002} {\bibfield
  {journal} {\bibinfo  {journal} {Phys. Rev. Applied}\ }\textbf {\bibinfo
  {volume} {2}},\ \bibinfo {pages} {054002} (\bibinfo {year}
  {2014})}\BibitemShut {NoStop}%
\bibitem [{\citenamefont {Lachance-Quirion}\ \emph {et~al.}(2020)\citenamefont
  {Lachance-Quirion}, \citenamefont {Piotr~Wolski}, \citenamefont {Tabuchi},
  \citenamefont {Kono}, \citenamefont {Usami},\ and\ \citenamefont
  {Nakamura}}]{magnonqubit}%
  \BibitemOpen
  \bibfield  {author} {\bibinfo {author} {\bibfnamefont {D.}~\bibnamefont
  {Lachance-Quirion}}, \bibinfo {author} {\bibfnamefont {S.}~\bibnamefont
  {Piotr~Wolski}}, \bibinfo {author} {\bibfnamefont {Y.}~\bibnamefont
  {Tabuchi}}, \bibinfo {author} {\bibfnamefont {S.}~\bibnamefont {Kono}},
  \bibinfo {author} {\bibfnamefont {K.}~\bibnamefont {Usami}}, \ and\ \bibinfo
  {author} {\bibfnamefont {Y.}~\bibnamefont {Nakamura}},\ }\bibfield  {title}
  {\emph {\bibinfo {title} {Entanglement-based single-shot detection of a
  single magnon with a superconducting qubit},\ }}\href@noop {} {\bibfield
  {journal} {\bibinfo  {journal} {Science}\ }\textbf {\bibinfo {volume}
  {367}},\ \bibinfo {pages} {425} (\bibinfo {year} {2020})}\BibitemShut
  {NoStop}%
\bibitem [{\citenamefont {Tabuchi}\ \emph {et~al.}(2015)\citenamefont
  {Tabuchi}, \citenamefont {Ishino}, \citenamefont {Noguchi}, \citenamefont
  {Ishikawa}, \citenamefont {Yamazaki}, \citenamefont {Usami},\ and\
  \citenamefont {Nakamura}}]{magnonqubit2}%
  \BibitemOpen
  \bibfield  {author} {\bibinfo {author} {\bibfnamefont {Y.}~\bibnamefont
  {Tabuchi}}, \bibinfo {author} {\bibfnamefont {S.}~\bibnamefont {Ishino}},
  \bibinfo {author} {\bibfnamefont {A.}~\bibnamefont {Noguchi}}, \bibinfo
  {author} {\bibfnamefont {T.}~\bibnamefont {Ishikawa}}, \bibinfo {author}
  {\bibfnamefont {R.}~\bibnamefont {Yamazaki}}, \bibinfo {author}
  {\bibfnamefont {K.}~\bibnamefont {Usami}}, \ and\ \bibinfo {author}
  {\bibfnamefont {Y.}~\bibnamefont {Nakamura}},\ }\bibfield  {title} {\emph
  {\bibinfo {title} {Coherent coupling between a ferromagnetic magnon and a
  superconducting qubit},\ }}\href@noop {} {\bibfield  {journal} {\bibinfo
  {journal} {Science}\ }\textbf {\bibinfo {volume} {349}},\ \bibinfo {pages}
  {405} (\bibinfo {year} {2015})}\BibitemShut {NoStop}%
\bibitem [{\citenamefont {Wallraff}\ \emph {et~al.}(2004)\citenamefont
  {Wallraff}, \citenamefont {Schuster}, \citenamefont {Blais}, \citenamefont
  {Frunzio},\ and\ \citenamefont {Huang}}]{superconducting}%
  \BibitemOpen
  \bibfield  {author} {\bibinfo {author} {\bibfnamefont {A.}~\bibnamefont
  {Wallraff}}, \bibinfo {author} {\bibfnamefont {D.~I.}\ \bibnamefont
  {Schuster}}, \bibinfo {author} {\bibfnamefont {A.}~\bibnamefont {Blais}},
  \bibinfo {author} {\bibfnamefont {L.}~\bibnamefont {Frunzio}}, \ and\
  \bibinfo {author} {\bibfnamefont {R.-S.}\ \bibnamefont {Huang}},\ }\bibfield
  {title} {\emph {\bibinfo {title} {Strong coupling of a single photon to a
  superconducting qubit using circuit quantum electrodynamics},\ }}\href@noop
  {} {\bibfield  {journal} {\bibinfo  {journal} {Nature}\ }\textbf {\bibinfo
  {volume} {431}},\ \bibinfo {pages} {162} (\bibinfo {year}
  {2004})}\BibitemShut {NoStop}%
\bibitem [{\citenamefont {Mi}\ \emph {et~al.}(2017)\citenamefont {Mi},
  \citenamefont {Cady}, \citenamefont {Zajac}, \citenamefont {Deelman},\ and\
  \citenamefont {Petta}}]{quantumdot}%
  \BibitemOpen
  \bibfield  {author} {\bibinfo {author} {\bibfnamefont {X.}~\bibnamefont
  {Mi}}, \bibinfo {author} {\bibfnamefont {J.~V.}\ \bibnamefont {Cady}},
  \bibinfo {author} {\bibfnamefont {D.~M.}\ \bibnamefont {Zajac}}, \bibinfo
  {author} {\bibfnamefont {P.~W.}\ \bibnamefont {Deelman}}, \ and\ \bibinfo
  {author} {\bibfnamefont {J.~R.}\ \bibnamefont {Petta}},\ }\bibfield  {title}
  {\emph {\bibinfo {title} {Strong coupling of a single electron in silicon to
  a microwave photon},\ }}\href@noop {} {\bibfield  {journal} {\bibinfo
  {journal} {Science}\ }\textbf {\bibinfo {volume} {355}},\ \bibinfo {pages}
  {156} (\bibinfo {year} {2017})}\BibitemShut {NoStop}%
\bibitem [{\citenamefont {Samkharadze}\ \emph {et~al.}(2018)\citenamefont
  {Samkharadze}, \citenamefont {Zheng}, \citenamefont {Kalhor}, \citenamefont
  {Brousse}, \citenamefont {Sammak}, \citenamefont {Mendes}, \citenamefont
  {Blais}, \citenamefont {Scappucci},\ and\ \citenamefont
  {Vandersypen}}]{quantumdot2}%
  \BibitemOpen
  \bibfield  {author} {\bibinfo {author} {\bibfnamefont {N.}~\bibnamefont
  {Samkharadze}}, \bibinfo {author} {\bibfnamefont {G.}~\bibnamefont {Zheng}},
  \bibinfo {author} {\bibfnamefont {N.}~\bibnamefont {Kalhor}}, \bibinfo
  {author} {\bibfnamefont {D.}~\bibnamefont {Brousse}}, \bibinfo {author}
  {\bibfnamefont {A.}~\bibnamefont {Sammak}}, \bibinfo {author} {\bibfnamefont
  {U.~C.}\ \bibnamefont {Mendes}}, \bibinfo {author} {\bibfnamefont
  {A.}~\bibnamefont {Blais}}, \bibinfo {author} {\bibfnamefont
  {G.}~\bibnamefont {Scappucci}}, \ and\ \bibinfo {author} {\bibfnamefont
  {L.~M.~K.}\ \bibnamefont {Vandersypen}},\ }\bibfield  {title} {\emph
  {\bibinfo {title} {Strong spin-photon coupling in silicon},\ }}\href@noop {}
  {\bibfield  {journal} {\bibinfo  {journal} {Science}\ }\textbf {\bibinfo
  {volume} {359}},\ \bibinfo {pages} {1123} (\bibinfo {year}
  {2018})}\BibitemShut {NoStop}%
\bibitem [{\citenamefont {Forn-D\'{\i}az}\ \emph {et~al.}(2010)\citenamefont
  {Forn-D\'{\i}az}, \citenamefont {Lisenfeld}, \citenamefont {Marcos},
  \citenamefont {Garc\'{\i}a-Ripoll}, \citenamefont {Solano}, \citenamefont
  {Harmans},\ and\ \citenamefont {Mooij}}]{strongcoupling1}%
  \BibitemOpen
  \bibfield  {author} {\bibinfo {author} {\bibfnamefont {P.}~\bibnamefont
  {Forn-D\'{\i}az}}, \bibinfo {author} {\bibfnamefont {J.}~\bibnamefont
  {Lisenfeld}}, \bibinfo {author} {\bibfnamefont {D.}~\bibnamefont {Marcos}},
  \bibinfo {author} {\bibfnamefont {J.~J.}\ \bibnamefont {Garc\'{\i}a-Ripoll}},
  \bibinfo {author} {\bibfnamefont {E.}~\bibnamefont {Solano}}, \bibinfo
  {author} {\bibfnamefont {C.~J. P.~M.}\ \bibnamefont {Harmans}}, \ and\
  \bibinfo {author} {\bibfnamefont {J.~E.}\ \bibnamefont {Mooij}},\ }\bibfield
  {title} {\emph {\bibinfo {title} {Observation of the bloch-siegert shift in a
  qubit-oscillator system in the ultrastrong coupling regime},\ }}\href
  {\doibase 10.1103/PhysRevLett.105.237001} {\bibfield  {journal} {\bibinfo
  {journal} {Phys. Rev. Lett.}\ }\textbf {\bibinfo {volume} {105}},\ \bibinfo
  {pages} {237001} (\bibinfo {year} {2010})}\BibitemShut {NoStop}%
\bibitem [{\citenamefont {Ridolfo}\ \emph {et~al.}(2013)\citenamefont
  {Ridolfo}, \citenamefont {Savasta},\ and\ \citenamefont
  {Hartmann}}]{strongcoupling2}%
  \BibitemOpen
  \bibfield  {author} {\bibinfo {author} {\bibfnamefont {A.}~\bibnamefont
  {Ridolfo}}, \bibinfo {author} {\bibfnamefont {S.}~\bibnamefont {Savasta}}, \
  and\ \bibinfo {author} {\bibfnamefont {M.~J.}\ \bibnamefont {Hartmann}},\
  }\bibfield  {title} {\emph {\bibinfo {title} {Nonclassical radiation from
  thermal cavities in the ultrastrong coupling regime},\ }}\href {\doibase
  10.1103/PhysRevLett.110.163601} {\bibfield  {journal} {\bibinfo  {journal}
  {Phys. Rev. Lett.}\ }\textbf {\bibinfo {volume} {110}},\ \bibinfo {pages}
  {163601} (\bibinfo {year} {2013})}\BibitemShut {NoStop}%
\bibitem [{\citenamefont {Zhao}\ \emph {et~al.}(2015)\citenamefont {Zhao},
  \citenamefont {Liu}, \citenamefont {Liu},\ and\ \citenamefont
  {Nori}}]{strongcoupling3}%
  \BibitemOpen
  \bibfield  {author} {\bibinfo {author} {\bibfnamefont {Y.-J.}\ \bibnamefont
  {Zhao}}, \bibinfo {author} {\bibfnamefont {Y.-L.}\ \bibnamefont {Liu}},
  \bibinfo {author} {\bibfnamefont {Y.-x.}\ \bibnamefont {Liu}}, \ and\
  \bibinfo {author} {\bibfnamefont {F.}~\bibnamefont {Nori}},\ }\bibfield
  {title} {\emph {\bibinfo {title} {Generating nonclassical photon states via
  longitudinal couplings between superconducting qubits and microwave fields},\
  }}\href {\doibase 10.1103/PhysRevA.91.053820} {\bibfield  {journal} {\bibinfo
   {journal} {Phys. Rev. A}\ }\textbf {\bibinfo {volume} {91}},\ \bibinfo
  {pages} {053820} (\bibinfo {year} {2015})}\BibitemShut {NoStop}%
\bibitem [{\citenamefont {Aspelmeyer}\ \emph {et~al.}(2014)\citenamefont
  {Aspelmeyer}, \citenamefont {Kippenberg},\ and\ \citenamefont
  {Marquardt}}]{optcavity}%
  \BibitemOpen
  \bibfield  {author} {\bibinfo {author} {\bibfnamefont {M.}~\bibnamefont
  {Aspelmeyer}}, \bibinfo {author} {\bibfnamefont {T.~J.}\ \bibnamefont
  {Kippenberg}}, \ and\ \bibinfo {author} {\bibfnamefont {F.}~\bibnamefont
  {Marquardt}},\ }\bibfield  {title} {\emph {\bibinfo {title} {Cavity
  optomechanics},\ }}\href {\doibase 10.1103/RevModPhys.86.1391} {\bibfield
  {journal} {\bibinfo  {journal} {Rev. Mod. Phys.}\ }\textbf {\bibinfo {volume}
  {86}},\ \bibinfo {pages} {1391} (\bibinfo {year} {2014})}\BibitemShut
  {NoStop}%
\bibitem [{\citenamefont {Zhang}\ \emph
  {et~al.}(2016{\natexlab{a}})\citenamefont {Zhang}, \citenamefont {Cheng},
  \citenamefont {Li},\ and\ \citenamefont {Zhou}}]{optcool}%
  \BibitemOpen
  \bibfield  {author} {\bibinfo {author} {\bibfnamefont {W.-Z.}\ \bibnamefont
  {Zhang}}, \bibinfo {author} {\bibfnamefont {J.}~\bibnamefont {Cheng}},
  \bibinfo {author} {\bibfnamefont {W.-D.}\ \bibnamefont {Li}}, \ and\ \bibinfo
  {author} {\bibfnamefont {L.}~\bibnamefont {Zhou}},\ }\bibfield  {title}
  {\emph {\bibinfo {title} {Optomechanical cooling in the non-markovian
  regime},\ }}\href {\doibase 10.1103/PhysRevA.93.063853} {\bibfield  {journal}
  {\bibinfo  {journal} {Phys. Rev. A}\ }\textbf {\bibinfo {volume} {93}},\
  \bibinfo {pages} {063853} (\bibinfo {year} {2016}{\natexlab{a}})}\BibitemShut
  {NoStop}%
\bibitem [{\citenamefont {Vitali}\ \emph {et~al.}(2007)\citenamefont {Vitali},
  \citenamefont {Gigan}, \citenamefont {Ferreira}, \citenamefont {B\"ohm},
  \citenamefont {Tombesi}, \citenamefont {Guerreiro}, \citenamefont {Vedral},
  \citenamefont {Zeilinger},\ and\ \citenamefont {Aspelmeyer}}]{optentang1}%
  \BibitemOpen
  \bibfield  {author} {\bibinfo {author} {\bibfnamefont {D.}~\bibnamefont
  {Vitali}}, \bibinfo {author} {\bibfnamefont {S.}~\bibnamefont {Gigan}},
  \bibinfo {author} {\bibfnamefont {A.}~\bibnamefont {Ferreira}}, \bibinfo
  {author} {\bibfnamefont {H.~R.}\ \bibnamefont {B\"ohm}}, \bibinfo {author}
  {\bibfnamefont {P.}~\bibnamefont {Tombesi}}, \bibinfo {author} {\bibfnamefont
  {A.}~\bibnamefont {Guerreiro}}, \bibinfo {author} {\bibfnamefont
  {V.}~\bibnamefont {Vedral}}, \bibinfo {author} {\bibfnamefont
  {A.}~\bibnamefont {Zeilinger}}, \ and\ \bibinfo {author} {\bibfnamefont
  {M.}~\bibnamefont {Aspelmeyer}},\ }\bibfield  {title} {\emph {\bibinfo
  {title} {Optomechanical entanglement between a movable mirror and a cavity
  field},\ }}\href {\doibase 10.1103/PhysRevLett.98.030405} {\bibfield
  {journal} {\bibinfo  {journal} {Phys. Rev. Lett.}\ }\textbf {\bibinfo
  {volume} {98}},\ \bibinfo {pages} {030405} (\bibinfo {year}
  {2007})}\BibitemShut {NoStop}%
\bibitem [{\citenamefont {Palomaki}\ \emph {et~al.}(2013)\citenamefont
  {Palomaki}, \citenamefont {Teufel}, \citenamefont {Marinkovi{\'c}},\ and\
  \citenamefont {L{\"o}schnauer}}]{optentang2}%
  \BibitemOpen
  \bibfield  {author} {\bibinfo {author} {\bibfnamefont {T.~A.}\ \bibnamefont
  {Palomaki}}, \bibinfo {author} {\bibfnamefont {J.~D.}\ \bibnamefont
  {Teufel}}, \bibinfo {author} {\bibfnamefont {I.}~\bibnamefont
  {Marinkovi{\'c}}}, \ and\ \bibinfo {author} {\bibfnamefont {C.}~\bibnamefont
  {L{\"o}schnauer}},\ }\bibfield  {title} {\emph {\bibinfo {title} {Entangling
  mechanical motion with microwave fields},\ }}\href@noop {} {\bibfield
  {journal} {\bibinfo  {journal} {Science}\ }\textbf {\bibinfo {volume}
  {342}},\ \bibinfo {pages} {710} (\bibinfo {year} {2013})}\BibitemShut
  {NoStop}%
\bibitem [{\citenamefont {Cheng}\ \emph {et~al.}(2019)\citenamefont {Cheng},
  \citenamefont {Liang}, \citenamefont {Zhang},\ and\ \citenamefont
  {Duan}}]{opttransfer}%
  \BibitemOpen
  \bibfield  {author} {\bibinfo {author} {\bibfnamefont {J.}~\bibnamefont
  {Cheng}}, \bibinfo {author} {\bibfnamefont {X.-T.}\ \bibnamefont {Liang}},
  \bibinfo {author} {\bibfnamefont {W.-Z.}\ \bibnamefont {Zhang}}, \ and\
  \bibinfo {author} {\bibfnamefont {X.}~\bibnamefont {Duan}},\ }\bibfield
  {title} {\emph {\bibinfo {title} {Optomechanical state transfer in the
  presence of non-markovian environments},\ }}\href {\doibase
  https://doi.org/10.1016/j.optcom.2018.08.079} {\bibfield  {journal} {\bibinfo
   {journal} {Opt. Commun.}\ }\textbf {\bibinfo {volume} {430}},\ \bibinfo
  {pages} {385 } (\bibinfo {year} {2019})}\BibitemShut {NoStop}%
\bibitem [{\citenamefont {Raimond}\ \emph {et~al.}(2001)\citenamefont
  {Raimond}, \citenamefont {Brune},\ and\ \citenamefont {Haroche}}]{qed}%
  \BibitemOpen
  \bibfield  {author} {\bibinfo {author} {\bibfnamefont {J.~M.}\ \bibnamefont
  {Raimond}}, \bibinfo {author} {\bibfnamefont {M.}~\bibnamefont {Brune}}, \
  and\ \bibinfo {author} {\bibfnamefont {S.}~\bibnamefont {Haroche}},\
  }\bibfield  {title} {\emph {\bibinfo {title} {Manipulating quantum
  entanglement with atoms and photons in a cavity},\ }}\href {\doibase
  10.1103/RevModPhys.73.565} {\bibfield  {journal} {\bibinfo  {journal} {Rev.
  Mod. Phys.}\ }\textbf {\bibinfo {volume} {73}},\ \bibinfo {pages} {565}
  (\bibinfo {year} {2001})}\BibitemShut {NoStop}%
\bibitem [{\citenamefont {Xiang}\ \emph {et~al.}(2013)\citenamefont {Xiang},
  \citenamefont {Ashhab}, \citenamefont {You},\ and\ \citenamefont
  {Nori}}]{qed2}%
  \BibitemOpen
  \bibfield  {author} {\bibinfo {author} {\bibfnamefont {Z.-L.}\ \bibnamefont
  {Xiang}}, \bibinfo {author} {\bibfnamefont {S.}~\bibnamefont {Ashhab}},
  \bibinfo {author} {\bibfnamefont {J.~Q.}\ \bibnamefont {You}}, \ and\
  \bibinfo {author} {\bibfnamefont {F.}~\bibnamefont {Nori}},\ }\bibfield
  {title} {\emph {\bibinfo {title} {Hybrid quantum circuits: Superconducting
  circuits interacting with other quantum systems},\ }}\href {\doibase
  10.1103/RevModPhys.85.623} {\bibfield  {journal} {\bibinfo  {journal} {Rev.
  Mod. Phys.}\ }\textbf {\bibinfo {volume} {85}},\ \bibinfo {pages} {623}
  (\bibinfo {year} {2013})}\BibitemShut {NoStop}%
\bibitem [{\citenamefont {Forn-D\'{\i}az}\ \emph {et~al.}(2019)\citenamefont
  {Forn-D\'{\i}az}, \citenamefont {Lamata}, \citenamefont {Rico}, \citenamefont
  {Kono},\ and\ \citenamefont {Solano}}]{qed3}%
  \BibitemOpen
  \bibfield  {author} {\bibinfo {author} {\bibfnamefont {P.}~\bibnamefont
  {Forn-D\'{\i}az}}, \bibinfo {author} {\bibfnamefont {L.}~\bibnamefont
  {Lamata}}, \bibinfo {author} {\bibfnamefont {E.}~\bibnamefont {Rico}},
  \bibinfo {author} {\bibfnamefont {J.}~\bibnamefont {Kono}}, \ and\ \bibinfo
  {author} {\bibfnamefont {E.}~\bibnamefont {Solano}},\ }\bibfield  {title}
  {\emph {\bibinfo {title} {Ultrastrong coupling regimes of light-matter
  interaction},\ }}\href {\doibase 10.1103/RevModPhys.91.025005} {\bibfield
  {journal} {\bibinfo  {journal} {Rev. Mod. Phys.}\ }\textbf {\bibinfo {volume}
  {91}},\ \bibinfo {pages} {025005} (\bibinfo {year} {2019})}\BibitemShut
  {NoStop}%
\bibitem [{\citenamefont {Reiserer}\ and\ \citenamefont {Rempe}(2015)}]{qed4}%
  \BibitemOpen
  \bibfield  {author} {\bibinfo {author} {\bibfnamefont {A.}~\bibnamefont
  {Reiserer}}\ and\ \bibinfo {author} {\bibfnamefont {G.}~\bibnamefont
  {Rempe}},\ }\bibfield  {title} {\emph {\bibinfo {title} {Cavity-based quantum
  networks with single atoms and optical photons},\ }}\href {\doibase
  10.1103/RevModPhys.87.1379} {\bibfield  {journal} {\bibinfo  {journal} {Rev.
  Mod. Phys.}\ }\textbf {\bibinfo {volume} {87}},\ \bibinfo {pages} {1379}
  (\bibinfo {year} {2015})}\BibitemShut {NoStop}%
\bibitem [{\citenamefont {Zhang}\ \emph
  {et~al.}(2016{\natexlab{b}})\citenamefont {Zhang}, \citenamefont {Zou},
  \citenamefont {Jiang},\ and\ \citenamefont {Tang}}]{magnoncavity}%
  \BibitemOpen
  \bibfield  {author} {\bibinfo {author} {\bibfnamefont {X.}~\bibnamefont
  {Zhang}}, \bibinfo {author} {\bibfnamefont {C.-L.}\ \bibnamefont {Zou}},
  \bibinfo {author} {\bibfnamefont {L.}~\bibnamefont {Jiang}}, \ and\ \bibinfo
  {author} {\bibfnamefont {H.}~\bibnamefont {Tang}},\ }\bibfield  {title}
  {\emph {\bibinfo {title} {Cavity magnonmechanics},\ }}\href@noop {}
  {\bibfield  {journal} {\bibinfo  {journal} {Sci. Adv.}\ }\textbf {\bibinfo
  {volume} {2}},\ \bibinfo {pages} {e1501286} (\bibinfo {year}
  {2016}{\natexlab{b}})}\BibitemShut {NoStop}%
\bibitem [{\citenamefont {Li}\ \emph {et~al.}(2018)\citenamefont {Li},
  \citenamefont {Zhu},\ and\ \citenamefont {Agarwal}}]{mppentang}%
  \BibitemOpen
  \bibfield  {author} {\bibinfo {author} {\bibfnamefont {J.}~\bibnamefont
  {Li}}, \bibinfo {author} {\bibfnamefont {S.-Y.}\ \bibnamefont {Zhu}}, \ and\
  \bibinfo {author} {\bibfnamefont {G.~S.}\ \bibnamefont {Agarwal}},\
  }\bibfield  {title} {\emph {\bibinfo {title} {Magnon-photon-phonon
  entanglement in cavity magnomechanics},\ }}\href {\doibase
  10.1103/PhysRevLett.121.203601} {\bibfield  {journal} {\bibinfo  {journal}
  {Phys. Rev. Lett.}\ }\textbf {\bibinfo {volume} {121}},\ \bibinfo {pages}
  {203601} (\bibinfo {year} {2018})}\BibitemShut {NoStop}%
\bibitem [{\citenamefont {Cirac}\ \emph {et~al.}(1997)\citenamefont {Cirac},
  \citenamefont {Zoller}, \citenamefont {Kimble},\ and\ \citenamefont
  {Mabuchi}}]{statetransfer1}%
  \BibitemOpen
  \bibfield  {author} {\bibinfo {author} {\bibfnamefont {J.~I.}\ \bibnamefont
  {Cirac}}, \bibinfo {author} {\bibfnamefont {P.}~\bibnamefont {Zoller}},
  \bibinfo {author} {\bibfnamefont {H.~J.}\ \bibnamefont {Kimble}}, \ and\
  \bibinfo {author} {\bibfnamefont {H.}~\bibnamefont {Mabuchi}},\ }\bibfield
  {title} {\emph {\bibinfo {title} {Quantum state transfer and entanglement
  distribution among distant nodes in a quantum network},\ }}\href {\doibase
  10.1103/PhysRevLett.78.3221} {\bibfield  {journal} {\bibinfo  {journal}
  {Phys. Rev. Lett.}\ }\textbf {\bibinfo {volume} {78}},\ \bibinfo {pages}
  {3221} (\bibinfo {year} {1997})}\BibitemShut {NoStop}%
\bibitem [{\citenamefont {Kimble}(2008)}]{statetransfer2}%
  \BibitemOpen
  \bibfield  {author} {\bibinfo {author} {\bibfnamefont {H.~J.}\ \bibnamefont
  {Kimble}},\ }\bibfield  {title} {\emph {\bibinfo {title} {The quantum
  internet},\ }}\href@noop {} {\bibfield  {journal} {\bibinfo  {journal}
  {Nature}\ }\textbf {\bibinfo {volume} {453}},\ \bibinfo {pages} {1023}
  (\bibinfo {year} {2008})}\BibitemShut {NoStop}%
\bibitem [{\citenamefont {Tian}\ and\ \citenamefont
  {Li}(2017)}]{stateconversion}%
  \BibitemOpen
  \bibfield  {author} {\bibinfo {author} {\bibfnamefont {L.}~\bibnamefont
  {Tian}}\ and\ \bibinfo {author} {\bibfnamefont {Z.}~\bibnamefont {Li}},\
  }\bibfield  {title} {\emph {\bibinfo {title} {Nonreciprocal quantum-state
  conversion between microwave and optical photons},\ }}\href {\doibase
  10.1103/PhysRevA.96.013808} {\bibfield  {journal} {\bibinfo  {journal} {Phys.
  Rev. A}\ }\textbf {\bibinfo {volume} {96}},\ \bibinfo {pages} {013808}
  (\bibinfo {year} {2017})}\BibitemShut {NoStop}%
\bibitem [{\citenamefont {Zhang}\ \emph {et~al.}(2003)\citenamefont {Zhang},
  \citenamefont {Peng},\ and\ \citenamefont {Braunstein}}]{statetransfer3}%
  \BibitemOpen
  \bibfield  {author} {\bibinfo {author} {\bibfnamefont {J.}~\bibnamefont
  {Zhang}}, \bibinfo {author} {\bibfnamefont {K.}~\bibnamefont {Peng}}, \ and\
  \bibinfo {author} {\bibfnamefont {S.~L.}\ \bibnamefont {Braunstein}},\
  }\bibfield  {title} {\emph {\bibinfo {title} {Quantum-state transfer from
  light to macroscopic oscillators},\ }}\href {\doibase
  10.1103/PhysRevA.68.013808} {\bibfield  {journal} {\bibinfo  {journal} {Phys.
  Rev. A}\ }\textbf {\bibinfo {volume} {68}},\ \bibinfo {pages} {013808}
  (\bibinfo {year} {2003})}\BibitemShut {NoStop}%
\bibitem [{\citenamefont {Genes}\ \emph {et~al.}(2008)\citenamefont {Genes},
  \citenamefont {Mari}, \citenamefont {Tombesi},\ and\ \citenamefont
  {Vitali}}]{statetransfer4}%
  \BibitemOpen
  \bibfield  {author} {\bibinfo {author} {\bibfnamefont {C.}~\bibnamefont
  {Genes}}, \bibinfo {author} {\bibfnamefont {A.}~\bibnamefont {Mari}},
  \bibinfo {author} {\bibfnamefont {P.}~\bibnamefont {Tombesi}}, \ and\
  \bibinfo {author} {\bibfnamefont {D.}~\bibnamefont {Vitali}},\ }\bibfield
  {title} {\emph {\bibinfo {title} {Robust entanglement of a micromechanical
  resonator with output optical fields},\ }}\href {\doibase
  10.1103/PhysRevA.78.032316} {\bibfield  {journal} {\bibinfo  {journal} {Phys.
  Rev. A}\ }\textbf {\bibinfo {volume} {78}},\ \bibinfo {pages} {032316}
  (\bibinfo {year} {2008})}\BibitemShut {NoStop}%
\bibitem [{\citenamefont {Schmidt}\ \emph {et~al.}(2012)\citenamefont
  {Schmidt}, \citenamefont {Ludwig},\ and\ \citenamefont
  {Marquardt}}]{stateprocess}%
  \BibitemOpen
  \bibfield  {author} {\bibinfo {author} {\bibfnamefont {M.}~\bibnamefont
  {Schmidt}}, \bibinfo {author} {\bibfnamefont {M.}~\bibnamefont {Ludwig}}, \
  and\ \bibinfo {author} {\bibfnamefont {F.}~\bibnamefont {Marquardt}},\
  }\bibfield  {title} {\emph {\bibinfo {title} {Optomechanical circuits for
  nanomechanical continuous variable quantum state processing},\ }}\href@noop
  {} {\bibfield  {journal} {\bibinfo  {journal} {New. J. Phys.}\ }\textbf
  {\bibinfo {volume} {14}},\ \bibinfo {pages} {125005} (\bibinfo {year}
  {2012})}\BibitemShut {NoStop}%
\bibitem [{\citenamefont {Macr\`{\i}}\ \emph {et~al.}(2018)\citenamefont
  {Macr\`{\i}}, \citenamefont {Nori},\ and\ \citenamefont {Kockum}}]{state1}%
  \BibitemOpen
  \bibfield  {author} {\bibinfo {author} {\bibfnamefont {V.}~\bibnamefont
  {Macr\`{\i}}}, \bibinfo {author} {\bibfnamefont {F.}~\bibnamefont {Nori}}, \
  and\ \bibinfo {author} {\bibfnamefont {A.~F.}\ \bibnamefont {Kockum}},\
  }\bibfield  {title} {\emph {\bibinfo {title} {Simple preparation of bell and
  greenberger-horne-zeilinger states using ultrastrong-coupling circuit qed},\
  }}\href {\doibase 10.1103/PhysRevA.98.062327} {\bibfield  {journal} {\bibinfo
   {journal} {Phys. Rev. A}\ }\textbf {\bibinfo {volume} {98}},\ \bibinfo
  {pages} {062327} (\bibinfo {year} {2018})}\BibitemShut {NoStop}%
\bibitem [{\citenamefont {Qi}\ and\ \citenamefont {Jing}(2020)}]{state2}%
  \BibitemOpen
  \bibfield  {author} {\bibinfo {author} {\bibfnamefont {S.}~\bibnamefont
  {Qi}}\ and\ \bibinfo {author} {\bibfnamefont {J.}~\bibnamefont {Jing}},\
  }\bibfield  {title} {\emph {\bibinfo {title} {Generating noon states in
  circuit qed using a multiphoton resonance in the presence of counter-rotating
  interactions},\ }}\href {\doibase 10.1103/PhysRevA.101.033809} {\bibfield
  {journal} {\bibinfo  {journal} {Phys. Rev. A}\ }\textbf {\bibinfo {volume}
  {101}},\ \bibinfo {pages} {033809} (\bibinfo {year} {2020})}\BibitemShut
  {NoStop}%
\bibitem [{\citenamefont {Garziano}\ \emph {et~al.}(2015)\citenamefont
  {Garziano}, \citenamefont {Stassi}, \citenamefont {Macr\`{\i}}, \citenamefont
  {Kockum}, \citenamefont {Savasta},\ and\ \citenamefont {Nori}}]{rabi2}%
  \BibitemOpen
  \bibfield  {author} {\bibinfo {author} {\bibfnamefont {L.}~\bibnamefont
  {Garziano}}, \bibinfo {author} {\bibfnamefont {R.}~\bibnamefont {Stassi}},
  \bibinfo {author} {\bibfnamefont {V.}~\bibnamefont {Macr\`{\i}}}, \bibinfo
  {author} {\bibfnamefont {A.~F.}\ \bibnamefont {Kockum}}, \bibinfo {author}
  {\bibfnamefont {S.}~\bibnamefont {Savasta}}, \ and\ \bibinfo {author}
  {\bibfnamefont {F.}~\bibnamefont {Nori}},\ }\bibfield  {title} {\emph
  {\bibinfo {title} {Multiphoton quantum rabi oscillations in ultrastrong
  cavity qed},\ }}\href {\doibase 10.1103/PhysRevA.92.063830} {\bibfield
  {journal} {\bibinfo  {journal} {Phys. Rev. A}\ }\textbf {\bibinfo {volume}
  {92}},\ \bibinfo {pages} {063830} (\bibinfo {year} {2015})}\BibitemShut
  {NoStop}%
\bibitem [{\citenamefont {Bender}\ \emph {et~al.}(2019)\citenamefont {Bender},
  \citenamefont {Kamra}, \citenamefont {Belzig},\ and\ \citenamefont
  {Duine}}]{magnoncoherence}%
  \BibitemOpen
  \bibfield  {author} {\bibinfo {author} {\bibfnamefont {S.~A.}\ \bibnamefont
  {Bender}}, \bibinfo {author} {\bibfnamefont {A.}~\bibnamefont {Kamra}},
  \bibinfo {author} {\bibfnamefont {W.}~\bibnamefont {Belzig}}, \ and\ \bibinfo
  {author} {\bibfnamefont {R.~A.}\ \bibnamefont {Duine}},\ }\bibfield  {title}
  {\emph {\bibinfo {title} {Spin current cross-correlations as a probe of
  magnon coherence},\ }}\href {\doibase 10.1103/PhysRevLett.122.187701}
  {\bibfield  {journal} {\bibinfo  {journal} {Phys. Rev. Lett.}\ }\textbf
  {\bibinfo {volume} {122}},\ \bibinfo {pages} {187701} (\bibinfo {year}
  {2019})}\BibitemShut {NoStop}%
\bibitem [{\citenamefont {de~Moraes~Neto}\ \emph {et~al.}(2016)\citenamefont
  {de~Moraes~Neto}, \citenamefont {Andrade}, \citenamefont {Montenegro},\ and\
  \citenamefont {Bose}}]{opttransfer2}%
  \BibitemOpen
  \bibfield  {author} {\bibinfo {author} {\bibfnamefont {G.~D.}\ \bibnamefont
  {de~Moraes~Neto}}, \bibinfo {author} {\bibfnamefont {F.~M.}\ \bibnamefont
  {Andrade}}, \bibinfo {author} {\bibfnamefont {V.}~\bibnamefont {Montenegro}},
  \ and\ \bibinfo {author} {\bibfnamefont {S.}~\bibnamefont {Bose}},\
  }\bibfield  {title} {\emph {\bibinfo {title} {Quantum state transfer in
  optomechanical arrays},\ }}\href {\doibase 10.1103/PhysRevA.93.062339}
  {\bibfield  {journal} {\bibinfo  {journal} {Phys. Rev. A}\ }\textbf {\bibinfo
  {volume} {93}},\ \bibinfo {pages} {062339} (\bibinfo {year}
  {2016})}\BibitemShut {NoStop}%
\bibitem [{\citenamefont {Singh}\ \emph {et~al.}(2012)\citenamefont {Singh},
  \citenamefont {Jing}, \citenamefont {Wright},\ and\ \citenamefont
  {Meystre}}]{ae1}%
  \BibitemOpen
  \bibfield  {author} {\bibinfo {author} {\bibfnamefont {S.}~\bibnamefont
  {Singh}}, \bibinfo {author} {\bibfnamefont {H.}~\bibnamefont {Jing}},
  \bibinfo {author} {\bibfnamefont {E.~M.}\ \bibnamefont {Wright}}, \ and\
  \bibinfo {author} {\bibfnamefont {P.}~\bibnamefont {Meystre}},\ }\bibfield
  {title} {\emph {\bibinfo {title} {Quantum-state transfer between a
  bose-einstein condensate and an optomechanical mirror},\ }}\href {\doibase
  10.1103/PhysRevA.86.021801} {\bibfield  {journal} {\bibinfo  {journal} {Phys.
  Rev. A}\ }\textbf {\bibinfo {volume} {86}},\ \bibinfo {pages} {021801}
  (\bibinfo {year} {2012})}\BibitemShut {NoStop}%
\bibitem [{\citenamefont {Vochezer}\ \emph {et~al.}(2018)\citenamefont
  {Vochezer}, \citenamefont {Kampschulte}, \citenamefont {Hammerer},\ and\
  \citenamefont {Treutlein}}]{ae2}%
  \BibitemOpen
  \bibfield  {author} {\bibinfo {author} {\bibfnamefont {A.}~\bibnamefont
  {Vochezer}}, \bibinfo {author} {\bibfnamefont {T.}~\bibnamefont
  {Kampschulte}}, \bibinfo {author} {\bibfnamefont {K.}~\bibnamefont
  {Hammerer}}, \ and\ \bibinfo {author} {\bibfnamefont {P.}~\bibnamefont
  {Treutlein}},\ }\bibfield  {title} {\emph {\bibinfo {title} {Light-mediated
  collective atomic motion in an optical lattice coupled to a membrane},\
  }}\href {\doibase 10.1103/PhysRevLett.120.073602} {\bibfield  {journal}
  {\bibinfo  {journal} {Phys. Rev. Lett.}\ }\textbf {\bibinfo {volume} {120}},\
  \bibinfo {pages} {073602} (\bibinfo {year} {2018})}\BibitemShut {NoStop}%
\bibitem [{\citenamefont {Xu}\ \emph {et~al.}(2019)\citenamefont {Xu},
  \citenamefont {Zhang},\ and\ \citenamefont {Liang}}]{ae3}%
  \BibitemOpen
  \bibfield  {author} {\bibinfo {author} {\bibfnamefont {X.}~\bibnamefont
  {Xu}}, \bibinfo {author} {\bibfnamefont {Z.}~\bibnamefont {Zhang}}, \ and\
  \bibinfo {author} {\bibfnamefont {Z.}~\bibnamefont {Liang}},\ }\bibfield
  {title} {\emph {\bibinfo {title} {Nonequilibrium quantum phase transition in
  a spinor quantum gas in a lattice coupled to a membrane},\ }}\href {\doibase
  10.1103/PhysRevA.100.053616} {\bibfield  {journal} {\bibinfo  {journal}
  {Phys. Rev. A}\ }\textbf {\bibinfo {volume} {100}},\ \bibinfo {pages}
  {053616} (\bibinfo {year} {2019})}\BibitemShut {NoStop}%
\bibitem [{\citenamefont {Christandl}\ \emph {et~al.}(2004)\citenamefont
  {Christandl}, \citenamefont {Datta}, \citenamefont {Ekert},\ and\
  \citenamefont {Landahl}}]{spin}%
  \BibitemOpen
  \bibfield  {author} {\bibinfo {author} {\bibfnamefont {M.}~\bibnamefont
  {Christandl}}, \bibinfo {author} {\bibfnamefont {N.}~\bibnamefont {Datta}},
  \bibinfo {author} {\bibfnamefont {A.}~\bibnamefont {Ekert}}, \ and\ \bibinfo
  {author} {\bibfnamefont {A.~J.}\ \bibnamefont {Landahl}},\ }\bibfield
  {title} {\emph {\bibinfo {title} {Perfect state transfer in quantum spin
  networks},\ }}\href {\doibase 10.1103/PhysRevLett.92.187902} {\bibfield
  {journal} {\bibinfo  {journal} {Phys. Rev. Lett.}\ }\textbf {\bibinfo
  {volume} {92}},\ \bibinfo {pages} {187902} (\bibinfo {year}
  {2004})}\BibitemShut {NoStop}%
\bibitem [{\citenamefont {Chang}\ and\ \citenamefont
  {Law}(2010)}]{nonmarkovmaster}%
  \BibitemOpen
  \bibfield  {author} {\bibinfo {author} {\bibfnamefont {K.~W.}\ \bibnamefont
  {Chang}}\ and\ \bibinfo {author} {\bibfnamefont {C.~K.}\ \bibnamefont
  {Law}},\ }\bibfield  {title} {\emph {\bibinfo {title} {Non-markovian master
  equation for a damped oscillator with time-varying parameters},\ }}\href
  {\doibase 10.1103/PhysRevA.81.052105} {\bibfield  {journal} {\bibinfo
  {journal} {Phys. Rev. A}\ }\textbf {\bibinfo {volume} {81}},\ \bibinfo
  {pages} {052105} (\bibinfo {year} {2010})}\BibitemShut {NoStop}%
\bibitem [{\citenamefont {Leggett}\ \emph {et~al.}(1987)\citenamefont
  {Leggett}, \citenamefont {Chakravarty}, \citenamefont {Dorsey}, \citenamefont
  {Fisher}, \citenamefont {Garg},\ and\ \citenamefont {Zwerger}}]{spectrum}%
  \BibitemOpen
  \bibfield  {author} {\bibinfo {author} {\bibfnamefont {A.~J.}\ \bibnamefont
  {Leggett}}, \bibinfo {author} {\bibfnamefont {S.}~\bibnamefont
  {Chakravarty}}, \bibinfo {author} {\bibfnamefont {A.~T.}\ \bibnamefont
  {Dorsey}}, \bibinfo {author} {\bibfnamefont {M.~P.~A.}\ \bibnamefont
  {Fisher}}, \bibinfo {author} {\bibfnamefont {A.}~\bibnamefont {Garg}}, \ and\
  \bibinfo {author} {\bibfnamefont {W.}~\bibnamefont {Zwerger}},\ }\bibfield
  {title} {\emph {\bibinfo {title} {Dynamics of the dissipative two-state
  system},\ }}\href {\doibase 10.1103/RevModPhys.59.1} {\bibfield  {journal}
  {\bibinfo  {journal} {Rev. Mod. Phys.}\ }\textbf {\bibinfo {volume} {59}},\
  \bibinfo {pages} {1} (\bibinfo {year} {1987})}\BibitemShut {NoStop}%
\bibitem [{\citenamefont {Gr{\"o}Blacher}\ \emph {et~al.}(2015)\citenamefont
  {Gr{\"o}Blacher}, \citenamefont {Trubarov}, \citenamefont {Prigge},
  \citenamefont {Cole}, \citenamefont {Aspelmeyer},\ and\ \citenamefont
  {Eisert}}]{spectrum2}%
  \BibitemOpen
  \bibfield  {author} {\bibinfo {author} {\bibfnamefont {S.}~\bibnamefont
  {Gr{\"o}Blacher}}, \bibinfo {author} {\bibfnamefont {A.}~\bibnamefont
  {Trubarov}}, \bibinfo {author} {\bibfnamefont {N.}~\bibnamefont {Prigge}},
  \bibinfo {author} {\bibfnamefont {G.~D.}\ \bibnamefont {Cole}}, \bibinfo
  {author} {\bibfnamefont {M.}~\bibnamefont {Aspelmeyer}}, \ and\ \bibinfo
  {author} {\bibfnamefont {J.}~\bibnamefont {Eisert}},\ }\bibfield  {title}
  {\emph {\bibinfo {title} {Observation of non-markovian micromechanical
  brownian motion},\ }}\href@noop {} {\bibfield  {journal} {\bibinfo  {journal}
  {Nat. Commun.}\ }\textbf {\bibinfo {volume} {6}},\ \bibinfo {pages} {7606}
  (\bibinfo {year} {2015})}\BibitemShut {NoStop}%
\bibitem [{\citenamefont {Wang}\ \emph {et~al.}(2018)\citenamefont {Wang},
  \citenamefont {Zhang}, \citenamefont {Zhang}, \citenamefont {Li},
  \citenamefont {Hu},\ and\ \citenamefont {You}}]{kerr1}%
  \BibitemOpen
  \bibfield  {author} {\bibinfo {author} {\bibfnamefont {Y.-P.}\ \bibnamefont
  {Wang}}, \bibinfo {author} {\bibfnamefont {G.-Q.}\ \bibnamefont {Zhang}},
  \bibinfo {author} {\bibfnamefont {D.}~\bibnamefont {Zhang}}, \bibinfo
  {author} {\bibfnamefont {T.-F.}\ \bibnamefont {Li}}, \bibinfo {author}
  {\bibfnamefont {C.-M.}\ \bibnamefont {Hu}}, \ and\ \bibinfo {author}
  {\bibfnamefont {J.~Q.}\ \bibnamefont {You}},\ }\bibfield  {title} {\emph
  {\bibinfo {title} {Bistability of cavity magnon polaritons},\ }}\href
  {\doibase 10.1103/PhysRevLett.120.057202} {\bibfield  {journal} {\bibinfo
  {journal} {Phys. Rev. Lett.}\ }\textbf {\bibinfo {volume} {120}},\ \bibinfo
  {pages} {057202} (\bibinfo {year} {2018})}\BibitemShut {NoStop}%
\bibitem [{\citenamefont {Wang}\ \emph {et~al.}(2016)\citenamefont {Wang},
  \citenamefont {Zhang}, \citenamefont {Zhang}, \citenamefont {Luo},
  \citenamefont {Xiong}, \citenamefont {Wang}, \citenamefont {Li},
  \citenamefont {Hu},\ and\ \citenamefont {You}}]{kerr2}%
  \BibitemOpen
  \bibfield  {author} {\bibinfo {author} {\bibfnamefont {Y.-P.}\ \bibnamefont
  {Wang}}, \bibinfo {author} {\bibfnamefont {G.-Q.}\ \bibnamefont {Zhang}},
  \bibinfo {author} {\bibfnamefont {D.}~\bibnamefont {Zhang}}, \bibinfo
  {author} {\bibfnamefont {X.-Q.}\ \bibnamefont {Luo}}, \bibinfo {author}
  {\bibfnamefont {W.}~\bibnamefont {Xiong}}, \bibinfo {author} {\bibfnamefont
  {S.-P.}\ \bibnamefont {Wang}}, \bibinfo {author} {\bibfnamefont {T.-F.}\
  \bibnamefont {Li}}, \bibinfo {author} {\bibfnamefont {C.-M.}\ \bibnamefont
  {Hu}}, \ and\ \bibinfo {author} {\bibfnamefont {J.~Q.}\ \bibnamefont {You}},\
  }\bibfield  {title} {\emph {\bibinfo {title} {Magnon kerr effect in a
  strongly coupled cavity-magnon system},\ }}\href {\doibase
  10.1103/PhysRevB.94.224410} {\bibfield  {journal} {\bibinfo  {journal} {Phys.
  Rev. B}\ }\textbf {\bibinfo {volume} {94}},\ \bibinfo {pages} {224410}
  (\bibinfo {year} {2016})}\BibitemShut {NoStop}%
\bibitem [{\citenamefont {Zhang}\ \emph {et~al.}(2019)\citenamefont {Zhang},
  \citenamefont {Wang},\ and\ \citenamefont {You}}]{kerr3}%
  \BibitemOpen
  \bibfield  {author} {\bibinfo {author} {\bibfnamefont {G.-Q.}\ \bibnamefont
  {Zhang}}, \bibinfo {author} {\bibfnamefont {Y.-P.}\ \bibnamefont {Wang}}, \
  and\ \bibinfo {author} {\bibfnamefont {J.~Q.}\ \bibnamefont {You}},\
  }\bibfield  {title} {\emph {\bibinfo {title} {Theory of the magnon kerr
  effect in cavity magnonics},\ }}\href@noop {} {\bibfield  {journal} {\bibinfo
   {journal} {Sci. China Phys. Mech. Astron.}\ }\textbf {\bibinfo {volume}
  {62}},\ \bibinfo {pages} {987511} (\bibinfo {year} {2019})}\BibitemShut
  {NoStop}%
\end{thebibliography}%

\end{document}